\documentclass[11pt]{article}
\usepackage{fancyhdr}
\usepackage{amsmath}
\usepackage{mathtools}
\usepackage{amsfonts}
\usepackage{amsmath}
\usepackage{amsbsy}
\usepackage{latexsym}
\usepackage[T1]{fontenc}
\usepackage[british,UKenglish,USenglish,english,american]{babel}
\usepackage{graphicx}
\newcommand{\RR}{\mathbb R}
\newcommand{\CC}{\mathbb C}

\newcommand{\ZZ}{{\mathbb Z}}

\renewcommand{\Im}{\mathop{\mathrm{Im}}}

\newcommand{\tr}{\mathop{\mathrm{tr}}}

 \newcommand{\beq}{\begin{equation}}
	\newcommand{\eeq}{\end{equation}}
\newcommand{\ba}{\begin{array}}
	\newcommand{\ea}{\end{array}}
\newcommand{\bea}{\begin{eqnarray}}
	\newcommand{\eea}{\end{eqnarray}}
\newcommand{\eps}{{\epsilon}}

\usepackage{graphicx}
\usepackage{xcolor}

\DeclareMathAlphabet{\mathpzc}{OT1}{pzc}{m}{it}

\begin{document}

\begin{center}
  {\bf $x$-periodic quasi one dimensional anomalous (rogue) waves \\ in multidimensional nonlinear Schr\"odinger equations: \\ fission, fusion, and recurrence}
\vskip 15pt
{\it F. Coppini $^{1,3,4}$ and P. M. Santini $^{1,2,5}$}

\vskip 20pt

{\it
$^1$ Dipartimento di Fisica, Universit\`a di Roma "La Sapienza", Piazz.le Aldo Moro 2, I-00185 Roma, Italy \\
$^2$ Istituto Nazionale di Fisica Nucleare (INFN), Sezione di Roma,  
Piazz.le Aldo Moro 2, I-00185 Roma, Italy \\
$^3$ Department of Mathematics, University at Buffalo. Mathematics Building UB North Campus. Buffalo NY, USA \\
}

\vskip 10pt
$^{4}$e-mail:  {\tt francesco.coppini@uniroma1.it,\\ fcoppini@buffalo.edu}\\
$^{5}$e-mail:  {\tt paolomaria.santini@fondazione.uniroma1.it, \\ paolo.santini@roma1.infn.it}\\
\bigskip
\vskip 10pt

{\today}

\end{center}

\begin{abstract}
  In a recent work we studied the first nonlinear stage of modulation instability (NLSMI) of $x$-periodic anomalous (rogue, freak, extreme) waves (AWs) of physically relevant multidimensional (generalizations  of the focusing) nonlinear Schr\"odinger (MNLS) equation, like the non integrable elliptic and hyperbolic nonlinear Schr\"odinger (NLS) equations in $d+1$ dimensions, $d=2,3$, in the quasi one dimensional (Q1D) regime in which the wavelength in the direction of propagation $x$ is small with respect to the wavelengths in the transversal directions. We showed that, at leading order, the first NLSMI is universal, independent of the particular MNLS model, and described by suitable adiabatic deformations of the quasi-homoclinic Akhmediev breather solution of NLS, in excellent agreement with numerical simulations. Varying the initial data, the first NLSMI consists in various combinations of the following basic processes: the AW growth from the unstable background, followed by fission in the transversal slowly varying directions, and the inverse process of fusion, followed by AW decay to the background. Here we study another important aspect of the theory: the recurrence of $x$-periodic AWs in the Q1D regime, showing the following. i) While the first NLSMI is essentially the same for all MNLS equations, there are $O(1)$ differences among the AW recurrences of the different MNLS equations. ii) Subsequent nonlinear stages of MI of the same MNLS equation exhibit, in general, more and more complicated combinations of fission and fusion processes, resulting in richer and richer choreographies. Since MNLS equations in the Q1D regime can be viewed as multidimensional perturbations of the integrable NLS equation, we use the recently developed finite gap perturbation theory of NLS AWs to give an analytic and quantitative description of the recurrence of Q1D AWs, in excellent agreement with numerical simulations. Due to the physical relevance of the MNLS equations considered in this work, and due to the universality of the processes discussed in this paper, it is plausible that they be observable in many fields of physics, like water waves, nonlinear optics, plasma physics, Bose-Einstein condensates, etc \dots 
\end{abstract}

\section{Introduction}

The self-focusing nonlinear Schr\"odinger (NLS) equation in $1+1$ dimensions
\beq\label{NLS}
i u_t +u_{xx}+2 |u|^2 u=0, \ \ u=u(x,t)\in\CC, \ \ x,t\in\RR,  
\eeq
is the simplest universal model in the description of the amplitude modulation of quasi monochromatic waves in weakly nonlinear self-focusing media; in particular, it is relevant in water waves \cite{Zakharov,AS}, in nonlinear optics \cite{Solli,Bortolozzo,PMContiADelRe}, in Langmuir waves in a plasma \cite{Malomed}, and in the theory of Bose-Einstein condensates \cite{Bludov,Pita}. It is well known that its homogeneous solution
\beq\label{background}
v_0(t)=\exp(2i t),
\eeq
describing Stokes waves \cite{Stokes} in a water wave context, a state of constant light intensity in nonlinear optics, and a state of constant boson density in a Bose-Einstein condensate, is unstable under the perturbation of waves with wave number $|k|<2$ and growth rate $\sigma(k)=|k|\sqrt{4-k^2}$ \cite{Bespalov,BF,Zakharov}, and this modulation instability (MI) is considered the main cause of the formation of anomalous (rogue, extreme, freak) waves (AWs) in nature \cite{KharifPeli3,Onorato2}. The integrability of \eqref{NLS}, based on the existence of the Zakharov - Shabat (ZS) zero-curvature representation \cite{ZakharovShabat}:
\begin{equation}
\label{eq:lp-x}
\vec\psi_x(\lambda,x,t)=\left[-i\lambda \sigma_3 +iU(x,t)\right]\vec\psi(\lambda,x,t),
\end{equation}
\begin{equation}
\label{eq:lp-t}
\vec\psi_t(\lambda,x,t)=V_{NLS}(\lambda,x,t)\vec\psi(\lambda,x,t),
\end{equation}
\beq\label{def_U_V}
\ba{l}
V_{NLS}=i\left(-2\lambda^2+U^2\right)\sigma_3+2i\lambda U-\sigma_3 U_x, \\
\sigma_3=\begin{pmatrix}  1 & 0 \\ 0 & -1 \end{pmatrix}, \ U = \begin{pmatrix}  0 & u\\ u^* & 0  \end{pmatrix}, \
\vec\psi= \left (\begin {array}{c} \psi_1 \\ \psi_2 \end {array}\right ),
\ea
\eeq
where $u^*$ is the complex conjugate of $u$, allows one to solve the NLS Cauchy problem for localized initial data \cite{ZakharovShabat}, using the inverse spectral transform introduced in \cite{GGKM}, and for space periodic initial data, using the finite gap method \cite{Novikov,Dubrovin,ItsMatveev,Lax,MKVM,Krichever}, and to construct large families of exact AW solutions, like the quasi-homoclinic Akhmediev breather (AB) \cite{Akhmed0}
\beq\label{Akhmed}
\ba{l}
Akh(x,t)=e^{2it}{\cal A}(x-x_0,t-t_0,\phi), \\
{\cal A}(x,t,\phi):=\frac{\cosh\left(\sigma t+2i\phi\right)+\sin(\phi)\cos(k x)}{\cosh(\sigma t)-\sin(\phi)\cos(k x)},\\
k=2\cos\phi, \ \ \sigma=\sigma(k)=|k|\sqrt{4-k^2}=2\sin(2\phi),
\ea
\eeq
space periodic and exponentially localized in time over the background \eqref{background}, its long wave limit, the Peregrine instanton \cite{Peregrine}, and their multi-mode generalizations \cite{ItsRybinSall,Dubard}. $x_0$ and $t_0$ in \eqref{Akhmed} are arbitrary real parameters associated with the space/time translation symmetries, and $(x_0,t_0)$ is the space-time point at which the AB \eqref{Akhmed} has its amplitude max: $|Akh(x_0,t_0)|=1+2\sin\phi$. The AB and the Peregrine instanton have been observed in many experiments; see, for instance: \cite{Yuen3,CHA_observP,KFFMDGA_observP,Kimmoun,Mussot,Pierangeli}.

In the NLS Cauchy problem for localized initial data over the background $v_0(t)$, slowly modulated periodic oscillations described by the elliptic solution of \eqref{NLS} play a relevant role in the longtime regime \cite{Biondini1,Biondini2}. The finite-gap method has been adapted in \cite{GS_FG_1,GS_FG_N} to solve at leading order and in terms of elementary functions the NLS Cauchy problem of AWs for periodic initial perturbations (of period $L$) of the background \eqref{background}, in the case of a finite number $N=\lfloor L/\pi\rfloor$ of unstable modes, showing the relevance of the $N$ breather solution of Akhmediev type in the description of the AW dynamics \cite{GS_FG_N}.

\begin{figure}[hhh!!!!!!!!!]
\begin{center}
\includegraphics[width=2.0cm,height=5.2cm]{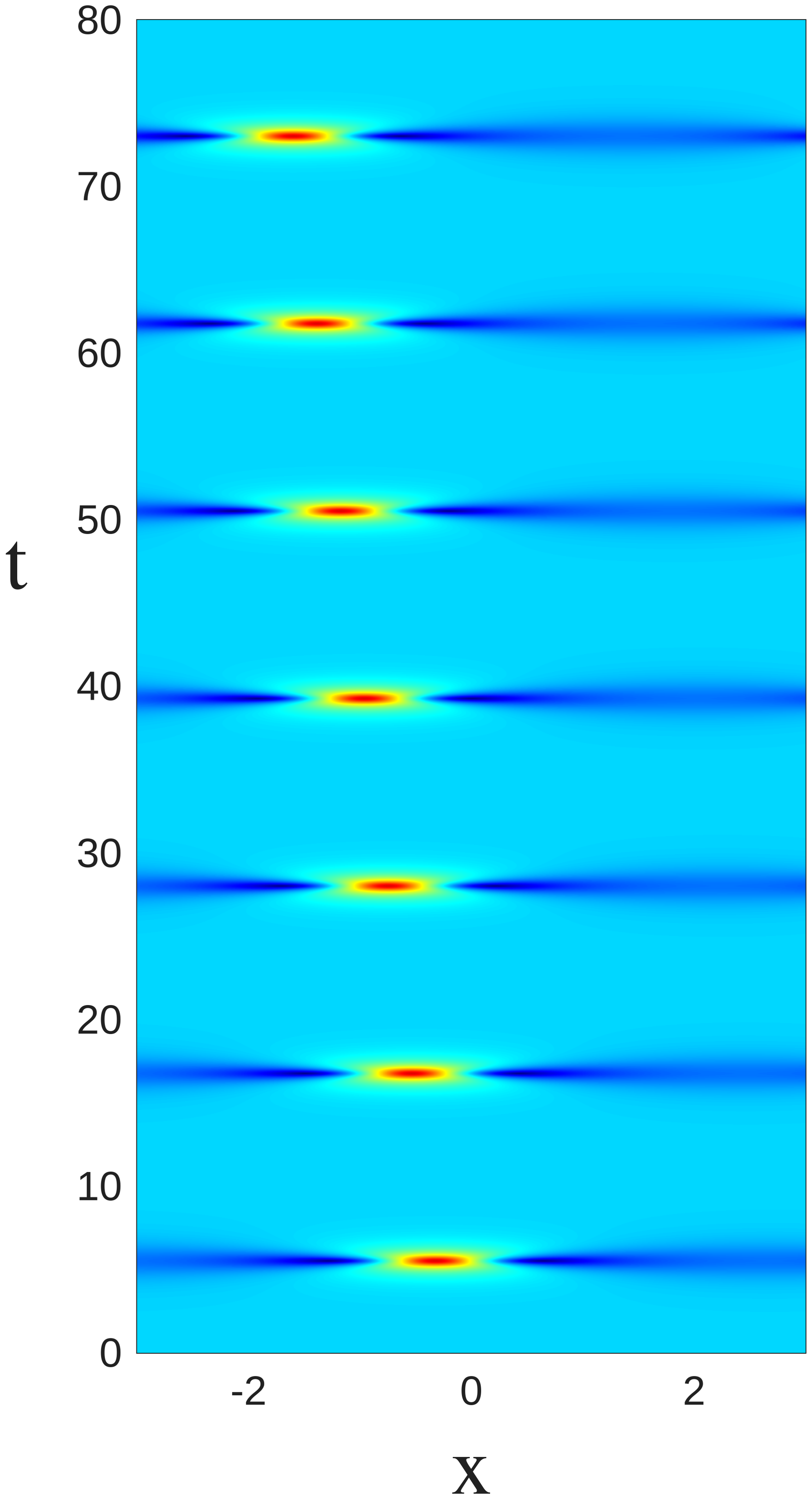} \ \ \includegraphics[width=9cm,height=5.2cm]{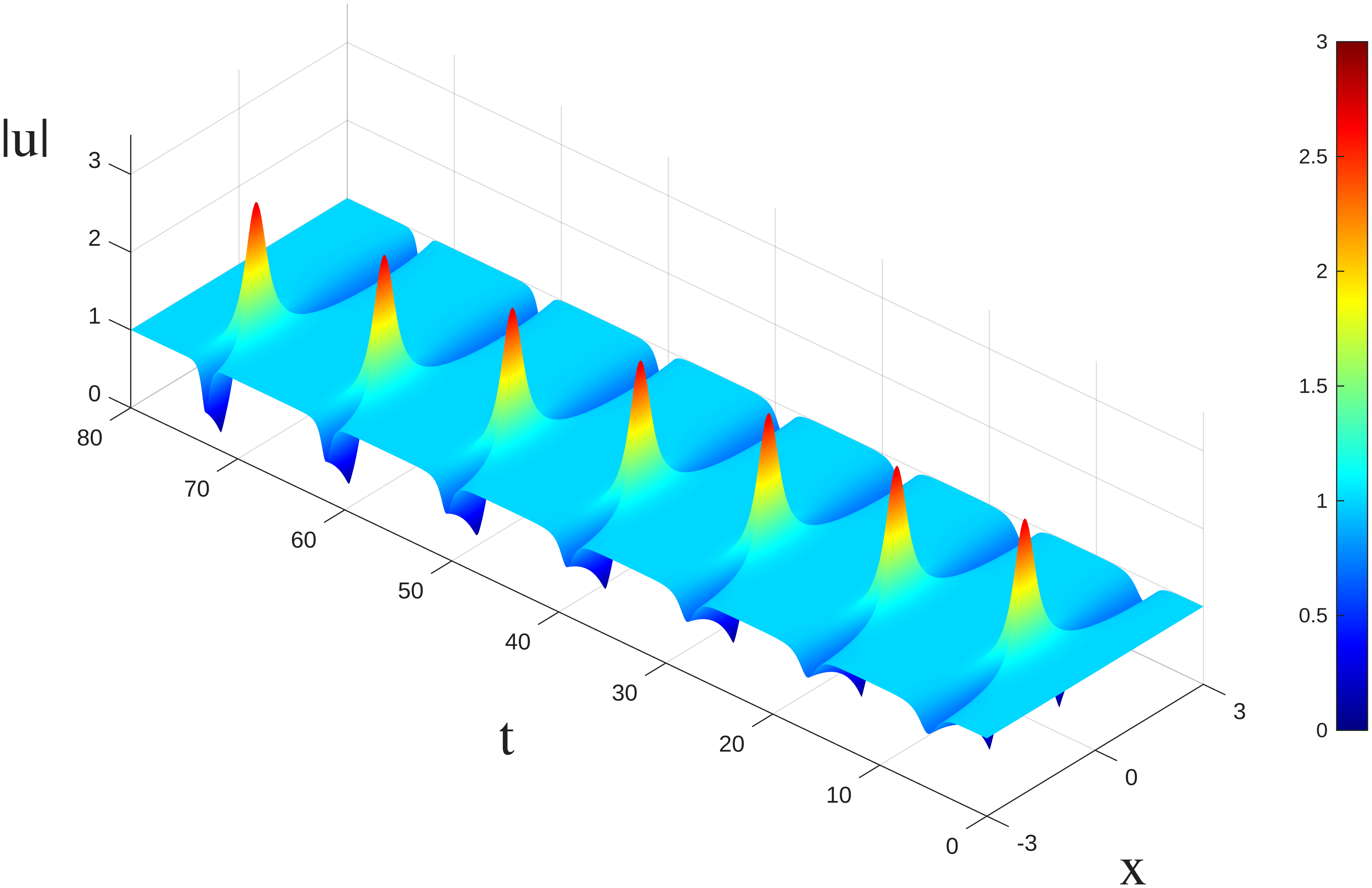}
\caption{The density plot (left) and the 3D plot (right) of the NLS numerical evolution of $|u(x,t)|$, for the generic initial condition \eqref{Cauchy1}, with $c_{-}=0.3+0.2i$, $c_+=0.5$, where the short axis is the $x$ axis, with $-L/2\le x\le L/2$, $0\le t \le 80$, $L=6$, $\epsilon=10^{-4}$, in extremely good quantitative agreement with (\ref{def_um}),(\ref{def_xm_tm}) \cite{GS_FG_1}.}\label{NLS_recurrence}
\end{center}
\end{figure}

In the simplest case of a single ($N=1$) unstable mode $k=2\pi/L$, for $\pi<L<2\pi$, the generic $O(\eps)$, $0<\eps\ll 1$ initial perturbation of $v_0(t)$
\beq\label{Cauchy1}
u(x,0)=1+\eps\left(c_+ e^{i k x}+c_{-}e^{-i k x}+\mbox{stable part of the Fourier series}\right),
\eeq
evolves into the following linear stage of MI (LSMI):
\beq\label{LSMI_NLS}
\ba{l}
u_{lin}(x,t)= e^{2it}\Big[1+\frac{2\eps}{\sigma}\Big(|\alpha_1 |e^{\sigma t+i\phi}\cos[k (x-x_1)]\\
+ |\alpha_0 |e^{-\sigma t-i\phi}\cos[k (x-{x}_0)]\Big)+O(\eps)\mbox{ oscillations}\Big], \ \ |t|=O(1)
\ea
\eeq
at leading order, where
\beq\label{def_x1_x0}
\ba{l}
x_1=\frac{\arg(\alpha_1)}{k}+\frac{L}{4}, \ \ x_0=\frac{-\arg(\alpha_0)}{k}+\frac{L}{4}, \ \ \mbox{mod }L,\\
\alpha_1 =e^{-i\phi}c^*_+-e^{i\phi}c_-, \ \ \ \alpha_0 =e^{i\phi}c^*_--e^{-i\phi}c_+, \\
\phi=\arccos(k/2), \ \ \sigma=\sigma(k)=2\sin(2\phi),
\ea
\eeq
followed by the first nonlinear stage of MI (NLSMI), at the logarithmically large time scale $O\!\left(\ln(1/\eps)\right)$, described at leading order by the AB
\beq\label{NLSMI1}
\ba{l}
u_1(x,t)={\cal A}\left(x-x_1,t-t_1,\phi\right)e^{2it+2i\phi}, \ \ |t-t_1|=O(1), \\
t_1=\frac{1}{\sigma}\log\!\left(\!\frac{\sigma^2}{2\eps |\alpha_1 |} \!\right),
\ea
\eeq
and then by a recurrence of linear and nonlinear stages of MI, such that the $m^{th}$ AW appearance is described at leading order by the AB \cite{GS_FG_1,GS_MAE}
\beq\label{def_um}
\ba{l}
u_m(x,t)={\cal A}\left(x-x_m,t-t_m,\phi\right)e^{2it+2i\phi_m}, \ \ |t-t_m|=O(1),
\ea
\eeq
where
\beq\label{def_xm_tm}
\ba{l}
x_m=x_1+(m-1)\Delta X, \ \ t_m=t_1 + (m-1)\Delta T, \ \ \phi_m=(2m-1)\phi, \\
\Delta X=\frac{\arg(\alpha_0\alpha_1)}{k}, \ \ \ \ \Delta T=\frac{1}{\sigma}\ln\!\left(\!\frac{\sigma^4}{4\eps^2 |\alpha_0\alpha_1 |} \!\right),
\ea
\eeq
giving a quantitative analytic description, in terms of elementary functions, of the AW recurrence observed in real and numerical experiments  \cite{Yuen1,Yuen3,Kimmoun,Mussot,Pierangeli}. $x_m$ and $t_m$ are respectively the position of the amplitude max of the AW and the time at which this max occurs, $\Delta X$ is the $x$-shift of the position of the AW between two consecutive appearances, and $\Delta T$ is the recurrence time (the time between two consecutive appearances). Therefore $t_1$ and $\Delta T$ are the $O(\log(1/\eps))$ characteristic times of MI and of the AW recurrence (see Figure \ref{NLS_recurrence}), a very nice and simple example, see \cite{CGS_FPUT}, of ideal Fermi-Pasta-Ulam-Tsingou type recurrence \cite{BI,Gallavotti}.

The recurrence \eqref{def_um},\eqref{def_xm_tm} for a single unstable mode can also be described by a simpler matched asymptotic expansion (MAE) approach \cite{GS_MAE}, not requiring the powerful tools of integrability, and therefore exportable to a much broader class of models, like the ones in this paper.

See also \cite{GS4} for a finite-gap model describing the numerical instabilities of the AB, and \cite{GS5} for the analytic study of the linear, nonlinear, and orbital instabilities of the AB within the NLS dynamics. See \cite{GS6} for the analytic study of the phase resonances in the AW recurrence, and see \cite{San}, \cite{CS_AL_rec} and \cite{CS_thesis} for the analytic study of the FPUT AW recurrence in other NLS type models: respectively the PT-symmetric NLS equation \cite{AM1}, the Ablowitz-Ladik lattice \cite{AL}, and  the massive Thirring model \cite{Thirring}.

The goal of this paper is to extend the above results on the NLS recurrence of AWs to multidimensional NLS-type equations in the quasi one dimensional (Q1D) regime in which the wavelength in the main direction of propagation is significantly smaller than the wavelengths in the transversal directions, and in the rest of this introduction we summarize the basic theoretical background needed to do it. 

\subsection{A finite gap description of the NLS AW recurrence}

Now we summarize some basic facts on the finite gap theory of NLS AWs developed in \cite{GS_FG_1,GS_FG_N}. Let $\Psi(\lambda,x,t)$ be a fundamental matrix solution of \eqref{eq:lp-x},\eqref{eq:lp-t},\eqref{def_U_V} for $x$-periodic solutions of period $L$, then the monodromy matrix
\beq\label{def_monodromy}
T(\lambda,t)=\Psi(\lambda,L,t)\Psi^{-1}(\lambda,0,t)
\eeq
is an entire function of $\lambda$, its  eigenvalues and eigenvectors are defined on a two-sheeted covering of the $\lambda$-plane, and this Riemann surface $\Gamma$ is a constant of motion.  The eigenvectors of $T(\lambda,t)$ are the Bloch eigenfunctions $\vec\psi_B(\gamma,x+L,t) =e^{iLp(\gamma)} \vec\psi_B(\gamma,x,t), \ \ \gamma \in \Gamma$, and the main spectrum is exactly the projection of the set $\{\gamma\in\Gamma,\Im p(\gamma)=0 \}$ to the $\lambda$-plane. The end points of the spectrum are the branch points and the double points (obtained merging pairs of branch points) of $\Gamma$, at which $e^{iLp(\gamma)}=\pm 1$, or, equivalently, $\tr T(\lambda)=\pm 2$. Due to the reality symmetry, if $\lambda$ belongs to the spectrum, also $\lambda^*$ does it.  
For the background $v_0(t)$,  the curve $\Gamma_0$ is rational and points $(\lambda,\mu)\in\Gamma_0$ are defined by equation: $\mu^2=\lambda^2+1$; the corresponding monodromy matrix $T_0$ is such that $\tr T_0(\lambda,t) = 2 \cos (\mu L)$, defining the branch points $(\pm i,0)$ and the double points $(\pm\lambda_n,\mu_n)=(\pm\sqrt{(n\pi/L)^2-1},n\pi/L)$, $n\in\ZZ,~n\ne 0$. Near $\lambda_n$:
\begin{equation}
\label{eq:prop4.1}
\tr T_0(\lambda) = (-1)^n \left[2 -\frac{\lambda_n^2 L^4}{\pi^2 n^2}(\lambda-\lambda_n)^2+O((\lambda-\lambda_n)^4)    \right].
\end{equation}

\begin{figure}[h!!!!!!!!!!!!!!!!!!!!!!!!!!!!!!]
	\centering
 \includegraphics[width=8cm]{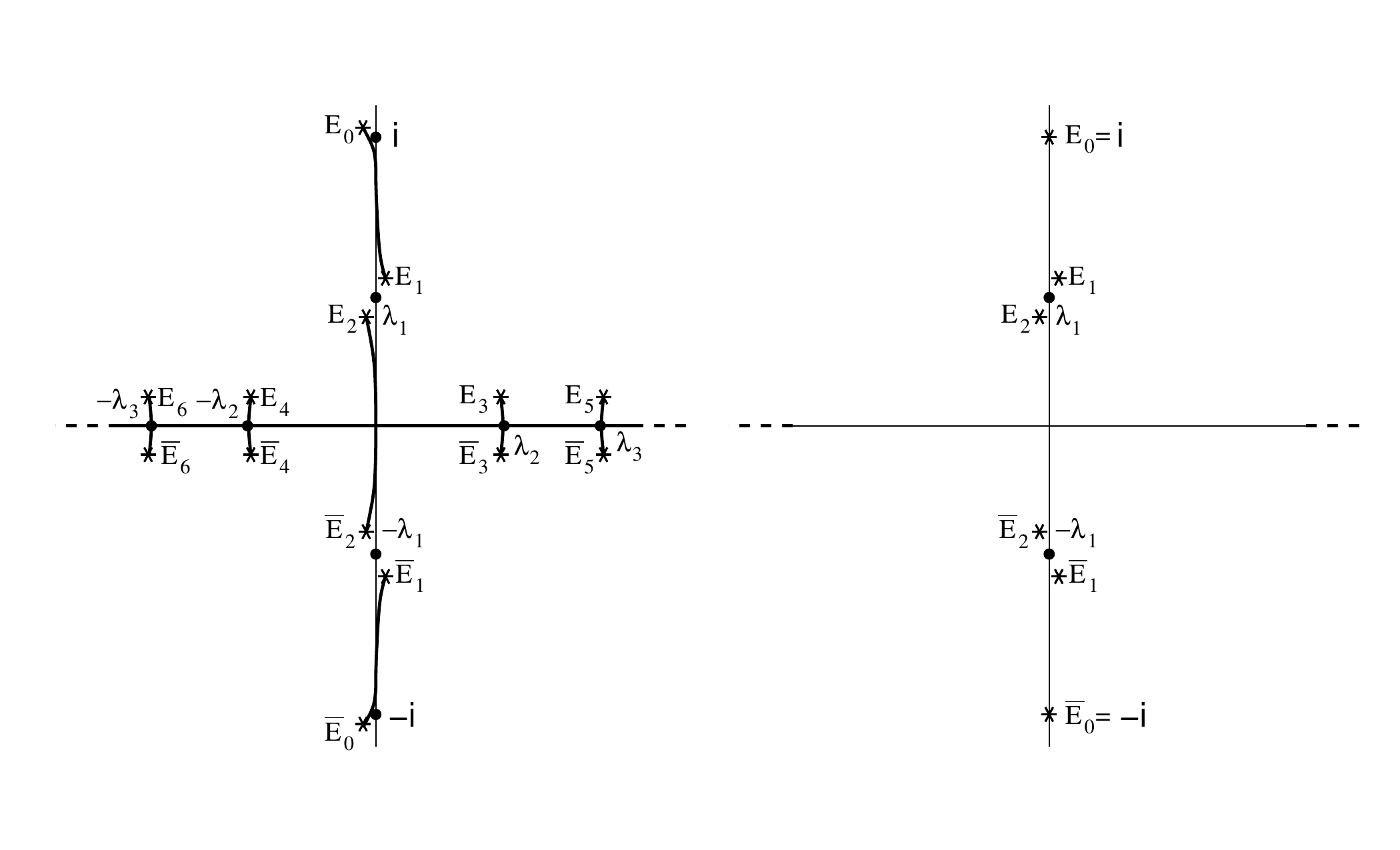}\\
  \caption{The infinitely many spines near the real axis of the left picture are erased in the finite gap approximation introduced in \cite{GS_FG_1,GS_FG_N}, obtaining the right picture.} \label{spectrum}
\end{figure}

Under the initial perturbation \eqref{Cauchy1} of the background, for $\eps\ll 1$, containing only the unstable mode $k=2\pi/L$ ($\pi<L <2\pi$), the branch points $\pm\lambda_0=\pm i$ become $E^{\pm}_0(Y)= \pm i+O(\eps^2)$, and the double point
\beq
\ba{l}
\lambda_1=i \sqrt{1-(\pi/L)^2}=i\sin\phi
\ea
\eeq
splits into the pair of square root branch points
\beq
\label{eq:bps2}
E_l =\lambda_1 +(-1)^l\frac{\epsilon}{2\lambda_1}\sqrt{\alpha_0\alpha_1}+O(\epsilon^2), \ \ l=1,2,
\eeq
generating the $O(\eps)$ gap centered at $\lambda_1$:
\beq\label{def_Gap0}
\mbox{Gap}_0=E_1-E_2=-\frac{\eps}{\lambda_1}\sqrt{\alpha_0\alpha_1}+O(\eps^2),
\eeq
together with its complex conjugate, and the trace of the monodromy matrix at $\lambda_1$ can be expressed at leading order in terms of the unstable gap through the following basic formula:
\beq\label{tr_gap}
\tr\left(T(\lambda_1,0) \right)=-2+\frac{\lambda_1^2 L^4}{4\pi^2}\mbox{Gap}_0^2.
\eeq
The infinitely many $O(\eps)$ spines generated by the splitting of the real double points $\pm \lambda_n, \ n\ge 2$, are closed, in the finite gap approximation introduced in \cite{GS_FG_1,GS_FG_N}, since they correspond to the stable modes, and give a negligeable contribution to the evolution of the initial data (see Figure \ref{spectrum}).

We shall also use the transition matrix
\beq\label{def_transition}
\hat T(\lambda,x,x',t)=\Psi(\lambda,x,t)\Psi^{-1}(\lambda,x',t)
\eeq
and, in particular, that corresponding to the AB \eqref{Akhmed} constructed in \cite{CGS_NLS_pert}
\begin{equation}\label{dressed_trans_matrix}
  \ba{l}
  \hat T(\lambda_1,x+L_x ,x,t) = e^{it\sigma_3}\tilde T(\lambda_1,x+L_x ,x,t)e^{-it\sigma_3}, \\
  \tilde T(\lambda_1,x+L_x ,x,t)=- I +i\frac{\pi}{2}\frac{\sin^2\!\phi}{\cos\phi}\frac{Q^*(x,t)}{\mbox{den}^2\!(x,t)}, 
\ea
\eeq
where
\begin{equation}\label{def_q}
  \ba{l}
  \mbox{den}(x,t) = \cosh\left(\sigma (t-t_0)\right)-\sin(\phi)\cos\left(k (x-x_0)\right), \\
  Q=\begin{pmatrix} - q_2 q_1 &  - q_2 q_2  \\ 
  q_1 q_1 &  q_1 q_2  \end{pmatrix}, \\
q_1=\sin\!\left(\frac{k(x-x_0)-\phi + i\sigma(t-t_0 )}{2}\right), \ \
q_2=-2i\cos\!\left(\frac{k(x-x_0)+\phi + i\sigma(t-t_0 )}{2}\right).
\ea
\end{equation}

\subsection{Perturbation theory of AWs}

To explain quantitatively the $O(1)$ effects of a small loss or gain on the NLS AW recurrence observed in real and numerical experiments \cite{Kimmoun,Soto}, a finite gap perturbation theory of AWs has been successfully constructed in \cite{CGS_NLS_pert} (see also \cite{CGS_Ency}), and here we briefly summarize it. For the perturbed NLS equation in $1+1$ dimensions
\beq\label{pert_NLS}
i u_t+u_{xx}+2|u|^2u=\nu\, p[u], \ \ |\nu |\ll 1,
\eeq
the trace of the monodromy matrix is no longer a constant of motion, and its variation in a time interval $\Delta t$ is given at leading order by \cite{CGS_NLS_pert}
\beq\label{variation_trT1}
\ba{l}
\Delta \tr {T} (\lambda,t)=\nu \!\!\int\limits_{\Delta t}\!\! dt\! \int\limits_0^{L}\!\! dx \left[{\hat T}_{21} (\lambda,x+L_x, x,t) p[u(x,t)]-\right.\\ \left. {\hat T}_{12} (\lambda,x+L_x, x,t) p^*[u(x,t)]\right], \\
\ea
\eeq
where $\hat T_{ij}$ is the $(i,j)$ component of the transition matrix. Consider a time interval $\Delta t$ containing the NLSMI described at leading order by the AB
\beq\label{um_pert}
u_m(x,t)={\cal A}\left(x-\tilde x_m,t-\tilde t_m \right)e^{2it+2i\phi_m}, \ \ |t-\tilde t_m|=O(1);
\eeq
since the off diagonal part of the transition matrix \eqref{dressed_trans_matrix} of \eqref{um_pert} is exponentially localized in time, the integral over the $x$-period is exponentially small during the LSMI preceeding the NLSMI described by \eqref{um_pert}. It follows that the variation of the trace of the monodromy matrix is negligeable during the LSMI preceeding the NLSMI described by \eqref{um_pert}, and is concentrated on the $O(1)$ time interval in which the AW \eqref{um_pert} appears. Then the integral \eqref{variation_trT1} can be well approximated at leading order by the integral involving \eqref{dressed_trans_matrix} and \eqref{um_pert} over the whole line $t\in\RR$:  
\beq\label{trace_variation_NLSMIm1}
\ba{l}
\Delta\!\tr T (\lambda)= \nu  J_m(\lambda), \\
J_m(\lambda)=\int\limits_{-\infty}^{\infty}\!\!\! dt \int\limits_{\RR}\!\! dx \Big[{\hat T}^{(m)}_{21}(\lambda,x'+L, x',t')\,p[e^{2it}{\cal A}(x',t')]\\ -{\hat T}^{(m)}_{12}(\lambda,x'+L, x',t')\,p^*[e^{2it}{\cal A}(x',t')]\Big], \\
x'=x-\tilde x_m, \ \ t'=t-\tilde t_m.
\ea
\eeq

Evaluating equation \eqref{trace_variation_NLSMIm1} at $\lambda=\lambda_1$, and using the relation \eqref{tr_gap} between $\tr T$ and the unstable gap, one infers that also the unstable gap is essentially constant during the LSMI preceeding the NLSMI described by \eqref{um_pert}, and its variation is concentrated during the $O(1)$ time interval in which the AW \eqref{um_pert} appears. Then the motion of the gap is essentially discrete, looking like the discrete motion of the hand of a clock \cite{CGS_NLS_pert}, and its discrete variation is described by the formula  
\beq\label{variation_gap} 
\ba{l}
\mbox{Gap}_m^2-\mbox{Gap}_{m-1}^2=-\nu \frac{4\cos^4\phi}{\pi^2 \sin^2\phi}J_m(\lambda_1) , \ \ m\ge 1,\\
\mbox{Gap}_0=-\frac{\eps}{\lambda_1}\sqrt{\alpha_0\alpha_1}.
\ea
\eeq
It is convenient to introduce the sequence $\{Q_m\}_{m\in\ZZ}$ of $O(1)$ elements
\beq\label{def_Qm}
Q_m=\frac{\lambda_1^2}{\eps^2}\mbox{Gap}^2_m,
\eeq
implying the recurrence
\beq\label{variation_Qm}
\ba{l}
Q_{m}=Q_{m-1} +\frac{\nu}{\eps^2}\frac{4\cos^4\!\phi}{\pi^2}J_m(\lambda_1), \ \ m\ge 1,\\
Q_0=\alpha_0\alpha_1.
\ea
\eeq

The other important formula of the recurrence establishes the connection between  two subsequent nonlinear stages of MI, described by $u_{m}$ and $u_{m+1}$, and the gap $\mbox{Gap}_m$ (or, better, $Q_m$, via \eqref{def_Qm}), constant during the logarithmically large time interval connecting these nonlinear stages:
\beq\label{variation_xm_tm}
\ba{l}
\tilde t_{m+1}-\tilde t_{m}=\frac{1}{\sigma}\ln\!\left(\frac{\sigma^4}{4\eps^2 |Q_m |} \right), \ \ \tilde t_1=t_1=\frac{1}{\sigma}\log\left(\frac{\sigma^2}{2\epsilon|\alpha_1|} \right), \\
\tilde x_{m+1}-\tilde x_{m}=\frac{\arg(Q_m)}{k}, \ \ \tilde x_1=x_1=\frac{\arg\alpha_1}{k_1}+\frac{L}{4}, \ \ \mbox{mod }L.
\ea
\eeq

If the recurrence time scale is much smaller than the perturbation time scale: $\ln(1/\eps)\ll 1/\nu$, equation \eqref{pert_NLS} remains a perturbation of NLS for a number of recurrences, and the $m^{th}$ AW appearance is described at leading order by the AB \eqref{um_pert} whose space-time parameters are obtained algorithmically from the recurrence \eqref{variation_Qm},\eqref{variation_xm_tm}. We observe that the small parameters $\eps,\nu$ appear in the recurrence \eqref{variation_Qm},\eqref{variation_xm_tm} through the ratio $\nu/\eps^2$. It follows that, no matter how small is the perturbation of NLS, if $\nu/\eps^2=O(1)$, the AW recurrence of the perturbed NLS equation \eqref{pert_NLS} shows $O(|J_m|)$ differences with respect to the AW recurrence of the unperturbed NLS equation \cite{CGS_NLS_pert}! This result, essentially due to MI, shows important differences with respect to soliton perturbation theory, in which an $O(\nu)$ perturbation of the NLS equation gives rise to $O(\nu)$ changes in the dynamics \cite{Kaup}. The first NLSMI is essentially the same for all perturbed NLS equations \eqref{pert_NLS}, and the same as that of NLS,  while the space-time position of the AW during the subsequent appearances depends on the particular perturbation through the integral $J_m$ in \eqref{trace_variation_NLSMIm1}, and formulas \eqref{variation_Qm},\eqref{variation_xm_tm}.

This theory has been applied in \cite{CS_GL_LL} to the Ginsburg-Landau \cite{Newell_Whitehead} and Lugiato-Lefevre \cite{LL} models, viewed as perturbations of NLS, and in \cite{CS_AL_pert} to hamiltonian and non hamiltonian perturbations of the Ablowitz-Ladik lattice \cite{AL}.

\subsection{Quasi one dimensional AWs: fission and fusion in the first NLSMI}

Our systematic investigation of the analytic properties of AWs in multidimensional generalizations of the NLS equation, hereafter called MNLS equations, has recently begun, based on the common sense argument of the lighthouse: the navigation in a dark night near the coast is difficult/dangerous, and the presence, here and there, of lighthouses is a great help. When studying MNLS equations we have two lighthouses; the first one is associated with the existence of integrable Davey Stewartson (DS) equations \cite{DS,Shulman}, and in \cite{GS7} a finite gap theory for doubly periodic AWs of the integrable DS2 equation has been constructed, allowing one to solve the periodic Cauchy problem of AWs at leading order in terms of elementary functions, like in the NLS case; while in \cite{CGS3} the $N$-breather AW solution of Akhmediev type of the integrable DS1 and DS2 equations has been constructed, playing a key role in the description of the AW recurrence in multidimensions, and in the phenomenological study of AWs of integrable and non integrable MNLS equations \cite{CS_2+1_phen}. See also \cite{Biondini3,Biondini4,Liu1,Liu2,Ohta1,Ohta2,Tajiri} for other examples of exact solutions of integrable DS equations.

The second lighthouse, common to integrable and non integrable MNLS equations, is the quasi one dimensional (Q1D) regime in which the wavelength in the direction of propagation $x$ is small with respect to the wavelengths in the transversal directions
\beq\label{Q1D}
\frac{\lambda_x}{\lambda_y}=\frac{\lambda_x}{\lambda_z}=\dots =O(\delta), \ \ 0<\delta \ll 1,
\eeq
and in \cite{CS_Q1D_NLSMI1} we investigated the first NLSMI of the non integrable and physically relevant elliptic NLS (ENLS) and hyperbolic NLS (HNLS) equations in $2+1$ dimensions
\beq\label{ENLS_HNLS_2+1}
  \ba{l}
      iu_t+u_{xx}+b\, u_{yy}+2 |u|^2 u=0, \ \ b=\pm 1,
      \ea
    \eeq
respectively for $b=1$ and $b=-1$, relevant respectively in Kerr optical media \cite{Kelley} and in the study of surface waves in deep water \cite{Zakharov}, and of their $3+1$ dimensional generalizations
\beq\label{ENLS_HNLS_3+1}
\ba{l}
iu_t+u_{xx}+u_{yy}+u_{zz}+2 |u|^2 u=0, \ \ ENLS, \\
iu_t+u_{xx}-\left(u_{yy}\pm u_{zz}\right)+2 |u|^2 u=0, \ \ HNLS_{\pm}.
\ea
\eeq
But the results are also valid for integrable and non integrable families of $2+1$ dimensional DS-type equations \cite{Benney,DS,ABB}, in which the complex amplitude $u$ is coupled with the real mean flow potential. It is well known that the two and three dimensional focusing effects of the laplacian in the ENLS equations lead to the blow up in finite time \cite{Sulem,Berge} of smooth initial data, while there is no blow up for the HNLS equation in $2+1$ dimensions, due to the defocusing effects in the $y$ direction.

Following the MAE approach introduced in \cite{GS_MAE}, we considered in \cite{CS_Q1D_NLSMI1} the Cauchy problem of AWs for the ENLS and HNLS equations \eqref{ENLS_HNLS_2+1} with the following initial perturbation of the background, periodic in $x$ (with period $L_x$ corresponding to a single unstable mode: $\pi<L_x<2\pi$), and slowly varying and even functions of $y$:
\beq\label{Cauchy_data_Q1D}
  \ba{l}
u(x,Y,0)=1+\eps \Big(c_{+}(Y)e^{i k_x x}+c_{-}(Y)e^{-i k_x x}+\mbox{stable part of the Fourier} \\
x\mbox{-series}\Big), \ \ k_x=\frac{2\pi}{L_x}, \ \ c_{\pm}(-Y)=c_{\pm}(Y), \ \ Y=\delta y, \ \ \delta=\frac{\lambda_x}{\lambda_y}, \ \ \eps,\delta\ll 1.
\ea
\eeq

In the LSMI, for $|t|\le O(1)$, described by $u=e^{2it}(1+\eps w)$, $iw_t+w_{xx}+b\,\delta^2w_{YY}+2(w+w^*)=0$, the solution is the same for both equations \eqref{ENLS_HNLS_2+1}, up to $O(\delta^2)$ corrections, and coincides with the LSMI \cite{GS_FG_1,GS_FG_N,GS_MAE} of the NLS equation, treating the slow variable $Y$ as parameter:
\beq\label{LSMIa}
\ba{l}
u_{lin}(x,Y,t)= \Big(1+\frac{2\eps}{\sigma\!_x}\sum\limits_{m=0}^1|\alpha_m(Y) |e^{(-1)^{m+1}\left(\sigma\!_x t+i\phi\right)}\cos[k_x (x-x_m(Y))]\\
+ O(\eps)\mbox{ oscillations}\Big)e^{2it},
\ea
\eeq
where
\beq\label{LSMIb}
\ba{l}
x_m(Y)=(-1)^{m+1}\frac{\arg(\alpha_m(Y)}{k_x}+\frac{L_x}{4}, \ \ \mbox{mod }L_x, \ \ m=0,1,\\
\alpha_1(Y) =e^{-i\phi}c^*_+(Y)-e^{i\phi}c_{-}(Y),\quad \alpha_0(Y) =e^{i\phi}c^*_{-}(Y)-e^{-i\phi}c_+(Y), \\
\phi=\arccos(k_x/2), \ \ \sigma\!_x=\sigma(k_x)=2\sin(2\phi).
\ea
\eeq
The first nonlinear stages of MI preceeding and following the LSMI \eqref{LSMIa},\eqref{LSMIb} are described, at leading order, by the following adiabatic deformations of the AB \eqref{Akhmed}
\beq\label{u_m(Y)}
\ba{l}
u_m(x,Y,t)={\cal A}(x-x_m(Y),t-t_m(Y),\phi)e^{2it+2i\phi_m}, \ \phi_m=(2m-1)\phi, \\
t_m(Y)=(-1)^{m+1}\frac{1}{\sigma\!_x}\ln\!\left(\frac{\sigma\!_x^2}{2 \eps |\alpha_m(Y)|}\right),\ \ |t-t_m(Y)|=O(1), \ m=0,1;
\ea
\eeq
$m=0$ for the NLSMI preceeding \eqref{LSMIa} at negative times, and $m=1$ for the one following \eqref{LSMIa} at positive times (see Figure \ref{inverse_t_scatt}).
\begin{figure}[h!!!!]
  \centering
  \includegraphics[trim=0cm 1cm 0cm 0 ,width=8.5cm,height=2.5cm]{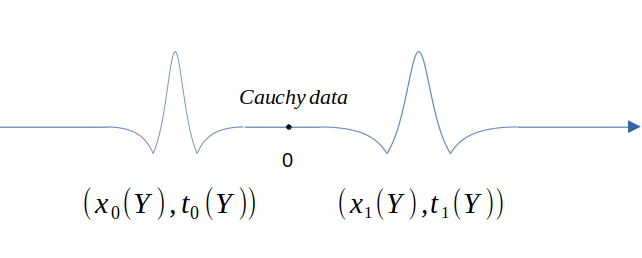}
  \caption{The schematic representation of the first nonlinear stages of MI for negative times $|t-t_0(Y)|=O(1)$ (left), and for positive times $|t-t_1(Y)|=O(1)$ (right), generated by the Cauchy data \eqref{Cauchy_data_Q1D}.} \label{inverse_t_scatt}
  \end{figure}

They give a universal description of the Q1D AW appearances, the same for all the MNLS equations, in terms of elementary functions, and in excellent agreement with numerical experiments \cite{CS_Q1D_NLSMI1}, if the MI and recurrence time scale $\ln(1/\eps)$ is much smaller than  the Q1D space-time scale $1/\delta$:
\beq
\ln\left(\frac{1}{\eps} \right)\ll \frac{1}{\delta}.
\eeq
This condition ensures that the multidimensional terms $\pm u_{yy}$ become relevant only after a number of recurrences, implying that blow up for the ENLS case, and  defocusing in the $y$ direction for the HNLS case, do not show up for a number of recurrences. 

Varying the initial data (the Fourier coefficients $c_{\pm}(Y)$ in \eqref{Cauchy_data_Q1D}), the first NLSMI \eqref{u_m(Y)} shows various combinations of the following basic and novel processes: a) the AW growth from the unstable background, followed by fission in the transversal slowly varying directions with infinite speed at fission time; b) the inverse process of fusion in the transversal slowly varying directions with infinite speed at fusion time, followed by AW decay to the background.

Some basic examples have been investigated in \cite{CS_Q1D_NLSMI1}, where the following results were also established. i) fission and fusion are critical processes showing similarities with other critical processes in nature: phase transitions of second kind and critical exponent $1/2$ \cite{Parisi}, and multidimensional wave breaking for dispersionless Kadomtsev-Petviashvili-type equations \cite{MS0,MS1,MS2}. ii) in $3+1$ dimensions with radial symmetry in the transversal slowly varying $(y,z)$ plane, fission corresponds to the generation of an opening smoke ring in the transversal plane centered on the $x$ axis, while if the symmetry is hyperbolic, fission corresponds to the splitting of an X wave into branches of hyperbola. iii) in the natural long wave limit, the Q1D AB reduces to the Q1D analogue of the Peregrine instanton. iv) the process of ``AW growth + fission'' is by no means restricted to the Q1D regime, extending to a finite region of the MI domain of the equation. The basic universal features of the first NLSMI are taken up again in Sections \ref{local},\ref{Examples}, in a more general discussion involving also the subsequent nonlinear stages of MI of the recurrence.
  
Indeed, in this paper we complete the results derived in \cite{CS_Q1D_NLSMI1} investigating another important property of $x$-periodic AW dynamics: the Q1D AW recurrence. In Section \ref{Recurrence} we derive the analytical formulas describing at leading order the AW recurrence in the Q1D regime for the ENLS and HNLS equations \eqref{ENLS_HNLS_2+1}. In Section \ref{local} we investigate the local and universal features of fission and fusion.  In Section \ref{Examples} we illustrate the theory on two basic examples: a doubly periodic Q1D AW, and an $x$-periodic AW localized over the background in the transversal direction. In Section \ref{numerics} we compare the analytical formulas of the paper with the numerical experiments.  In Section \ref{Conclusions} we present a reasoned summary of the main results of this paper, and we list some of the open problems we are presently investigating or we plan to investigate in the future.

Due to the physical relevance of the MNLS equations considered in this work, and due to the universality of the processes discussed in this paper, it is plausible that they be observable in natural phenomena in which AWs of multidimensional NLS equations are relevant, like in water waves, nonlinear optics, plasma physics, Bose-Einstein condensates, etc \dots .

\section{The recurrence of Q1D AWs of the ENLS and HNLS equations}\label{Recurrence}

When dealing with Q1D AWs of MNLS equations, one expects that suitable adiabatic multidimensional deformations of the quasi homoclinic AB \eqref{Akhmed} and of the Peregrine instanton be relevant, and this expectation is confirmed by the first NLSMI, described at leading order and in excellent agreement with numerical experiments, by the slowly varying analogue \eqref{u_m(Y)} of the AB \cite{CS_Q1D_NLSMI1}. In addition this representation is universal, the same for all MNLS equations, and again a  heuristic and rather naive argument can be used to justify it: ``in the Q1D regime all MNLS equations are close to NLS, and then close to each other''.  Now we show that this heuristic argument fails when investigating another important aspect of $x$-periodic AW dynamics: the recurrence of Q1D AWs, that turns out to be significantly different for different MNLS equations.

Since MNLS equations in the Q1D regime can be viewed as multidimensional perturbations of the integrable NLS equation, here we use the recently developed finite gap perturbation theory of NLS AWs \cite{CGS_NLS_pert} to construct the proper analytic finite gap model describing quantitatively and in terms of elementary functions the reasons of these differences, in excellent agreement with numerical experiments.

In the Q1D regime, the ENLS and HNLS equations \eqref{ENLS_HNLS_2+1} can be written as multidimensional perturbations of the NLS equation:
\beq\label{ENLS_HNLS_pert} 
\ba{l}
i u_t+u_{xx}+2|u|^2 u=-b\delta^2 u_{YY}, \ \ u=u(x,Y,t)\in\CC, \ \ Y=\delta y, \ \ \delta\ll 1,
\ea
\eeq
or, in matrix form:
\beq\label{matrix_form_2+1}
U_t=i\sigma_3(U_{xx}+2U^3)+ib \delta^2 \sigma_3 U_{YY}, \  \ b=\pm 1, \ \ U = \begin{pmatrix}  0 & u\\ u^* & 0  \end{pmatrix}.
\eeq

Equations \eqref{matrix_form_2+1} are the compatibility condition of the ZS spectral problem \eqref{eq:lp-x} in which $Y$ appears parametrically:
\beq\label{ZS_Y}
\Psi_x(\lambda,x,Y,t)=\left[-i\lambda \sigma_3 +iU(x,Y,t)\right]\Psi(\lambda,x,Y,t),
\eeq
and the nonlinear, nonlocal, and non integrable  time evolution
\beq\label{t-part}
\ba{l}
\Psi_t(\lambda,x,Y,t)=V_{NLS}(\lambda,x,Y,t)\Psi(\lambda,x,Y,t)\\
-b\delta^2\Psi(\lambda,x,Y,t)\int\limits_0^x\!\Psi^{-1}(\lambda,x',Y,t)\sigma_3U_{YY}(x',Y,t)\Psi(\lambda,x',Y,t)dx',
\ea
\eeq
where $V_{NLS}$ is defined in \eqref{def_U_V}, and $\Psi$ is any fundamental matrix solution. In general, this representation is useless, unless the nonlinear term in \eqref{t-part} is small, and a perturbation theory can be developed.

\subsection{t-evolution of the slowly varying unstable gap}

The Q1D deformations \eqref{u_m(Y)} of the AB \eqref{Akhmed} describing the first NLSMI for positive and negative times, together with the parametric dependence of the ZS spectral problem \eqref{ZS_Y} on the slow variable $Y$, suggest that the Riemann surface associated with the initial condition \eqref{Cauchy_data_Q1D} is that of NLS, but suitably depending parametrically on the slow variable $Y$, with the slowly varying unstable gap (see \eqref{def_Gap0}) 
\beq\label{Gap0(Y)}
\mbox{Gap}_0(Y)=-\frac{\eps}{\lambda_1}\sqrt{\alpha_0(Y)\alpha_1(Y)},
\eeq
where $\alpha_0(Y)$ and $\alpha_1(Y)$ are defined in \eqref{LSMIb}. Its time evolution according to the ENLS and the HNLS equations, and, in general, according to any MNLS equation viewed as multidimensional perturbations of NLS, can be described as follows.

Let
\beq\label{def_trans_Q1D}
\ba{l}
\hat T(\lambda,x,x',Y,t):=\Psi(\lambda,x,Y,t)\Psi^{-1}(\lambda,x',Y,t), \ \ 0\le x'< x\le L_x, \\
T(\lambda,Y,t):=\hat T(\lambda,L_x,0,Y,t)
\ea
\eeq
be respectively the slowly varying transition matrix and monodromy matrix of equations \eqref{ZS_Y},\eqref{t-part}. Omitting for the sake of simplicity their $\lambda,Y,t$ dependence, equations \eqref{ZS_Y},\eqref{t-part},\eqref{def_trans_Q1D} imply the following time evolution of the transition matrix:
\beq\label{hatT_evol}
\ba{l}
\hat T_t(x,x')=V_{NLS}\!\left(x\right)\hat T(x,x')-\hat T(x,x')V_{NLS}\!\left(x'\right)\\
-b\delta^2\int\limits_{x'}^x \hat T(x,x'')\sigma_3 U_{YY}(x'')\hat T(x'',x')dx''.
\ea
\eeq
Using the $L_x$-periodicity of $V_{NLS}$: $V_{NLS}(0)=V_{NLS}(L_x)$, and (\ref{def_trans_Q1D}b), in the limit $x\!\uparrow\! L_x, \ x'\!\downarrow\! 0$ one obtains the time evolution of the monodromy matrix $T(\lambda,Y,t)$:
\beq\label{T_evol}
\ba{l}
T_t=\left[V_{NLS}\left(0\right),T\right]
-b\delta^2\int\limits_{0}^{L_x}\!\hat T(L_x,x)\sigma_3 U_{YY}(x)\hat T(x,0)dx,
\ea
\eeq
where $[\cdot,\cdot ]$ is the commutator of two matrices. Then the trace of  \eqref{T_evol} yields
\beq\label{trT_evol}
\ba{l}
\left(\tr T \right)_t=-b\delta^2\int\limits_{0}^{L_x}\!\tr\!\left(\hat T(L_x,x)\sigma_3 U_{YY}(x)\hat T(x,0)\right)\! dx=\\
-b\delta^2\int\limits_{0}^{L_x}\!\tr\!\left({\hat T}^{-1}(x+L_x,L_x){\hat T}(x+L_x,L_x)\hat T(L_x,x)\sigma_3 U_{YY}(x)\hat T(x,0)\right)\! dx=\\
-b\delta^2\int\limits_{0}^{L_x}\!\tr\!\left({\hat T}(x+L_x,L_x)\hat T(L_x,x)\sigma_3 U_{YY}(x)\hat T(x,0){\hat T}^{-1}(x+L_x,L_x)\right)\! dx=\\
-b\delta^2\int\limits_{0}^{L_x}\!\tr\!\left({\hat T}(x+L_x,x)\sigma_3 U_{YY}(x)\right)\! dx,
\ea
\eeq
where, as in \cite{CGS_NLS_pert}, we have used the properties ${\hat T}(x,x'){\hat T}(x',x'')={\hat T}(x,x'')$ and ${\hat T}(x,x')={\hat T}(x+L_x,x'+L_x)$. Therefore the variation of $\tr T$ in a generic time interval $\Delta t$ is described by the slowly varying analogue of \eqref{variation_trT1}:
\beq\label{variation of trT}
\ba{l}
\Delta \tr {T} (\lambda, Y,t)=\!\! -b\delta^2 \!\!\int\limits_{\Delta t}\!\! dt\! \int\limits_0^{L_x}\!\! dx \Big\{{\hat T}_{21} (\lambda,x+L_x, x,Y,t) u_{YY}(x,Y,t)\\
-{\hat T}_{12} (\lambda,x+L_x, x,Y,t) u^*_{YY}(x,Y,t)\Big\}.
\ea
\eeq

Reasoning as in \cite{CGS_NLS_pert}, we consider an interval $\Delta t$ containing a single AW appearance described by $u_m$ in \eqref{u_m(Y)}. Since the off-diagonal part of its transition matrix \eqref{dressed_trans_matrix} is exponentially localized in time, the integral over the $x$-period is exponentially small in time during the LSMI preceeding the AW appearance, implying that $\tr {T} (\lambda, Y,t)$ is essentially constant during the LSMI, and its variation is concentrated during the AW appearance (see Figure \ref{trT_cos} of Section \ref{Examples}). Then the integral \eqref{variation of trT} over an interval $\Delta t$ containing the appearance described by \eqref{u_m(Y)} can be well approximated at leading order by the following integral over the whole line $t\in\RR$:  
\beq\label{trace_variation_NLSMIm}
\ba{l}
\Delta_m \!\tr T (\lambda,Y)= -b\delta^2 J_m(\lambda,Y), \\
\ea
\eeq
where
\beq\label{def_Jm(Y)}
\ba{l}
J_m(\lambda,Y)=\int\limits_{-\infty}^{\infty}\!\!\! dt \int\limits_0^{L_x}\!\! dx \Big[{\tilde T}^{(m)}_{21}(\lambda_1,x'+L_x, x',t')\,{\cal A}_{YY}(x',t')\\
-{\tilde T}^{(m)}_{12}(\lambda_1,x'+L_x, x',t')\,{\cal A}^*_{YY}(x',t')\Big], \\
x'=x-x_m(Y), \ \ t'=t-t_m(Y),
\ea
\eeq
where $\tilde T^{(m)}$, corresponding to \eqref{u_m(Y)}, is defined in \eqref{dressed_trans_matrix}, and $\tilde T^{(m)}_{ij}$ is its $(i,j)$ component. Evaluating equations \eqref{trace_variation_NLSMIm},\eqref{def_Jm(Y)} at $\lambda=\lambda_1$, and using the slowly varying analogue of \eqref{tr_gap}
\beq
\tr\left(T(\lambda_1,Y,t) \right)=-2+\frac{\lambda_1^2 L_x^4}{4\pi^2}\mbox{Gap}^2(Y,t),
\eeq
one infers that also the slowly varying unstable gap \eqref{Gap0(Y)} is essentially constant in time during the logarithmically large time interval in which the solution is essentially background, and its variation is concentrated during the $O(1)$ time interval in which the AW \eqref{u_m(Y)} appears. Then the evolution of the gap is essentially discrete, looking like the discrete motion of the hand of a clock, as in the $1+1$ dimensional NLS perturbation theory \cite{CGS_NLS_pert}, and is described by a sequence of slowly varying complex functions $\{\mbox{Gap}_m(Y)\}_{m\ge 0}$ generated at leading order by the recurrence  
\beq\label{variation_trT_gap} 
\ba{l}
\mbox{Gap}_m^2(Y)-\mbox{Gap}_{m-1}^2(Y)=-b\delta^2 \frac{4\cos^4\phi}{\pi^2 \sin^2\phi}J_m(\lambda_1,Y) , \ m\ge 1, \ \eta=\pm 1, \\
\mbox{Gap}_0(Y)=-\frac{\eps}{\lambda_1}\sqrt{\alpha_0(Y)\alpha_1(Y)}.
\ea
\eeq

\subsection{Explicit evaluation of $J_m(\lambda_1,Y)$}

It is remarkable that the double integral $J_m(\lambda_1,Y)$ in \eqref{def_Jm(Y)}, describing the discrete dynamics of the slowly varying unstable gap, can be evaluated explicitely as follows in terms of elementary functions:
\beq\label{Jm_final}
J_m(\lambda_1,Y)=\frac{4\pi^2\sigma\!_x^2}{k_x^6}\!\left[\frac{5}{3}\sigma\!_x \,\ddot t_m(Y)+i \, k_x\ddot x_m(Y)\right].
\eeq

To show it, we first observe that the $Y$ dependence of the integrand of $J_m(\lambda_1,Y)$ in \eqref{def_Jm(Y)} through the arguments $x'=x-x_m(Y)$ and $t'=t-t_m(Y)$ of $u_m$ and $\tilde T^{(m)}$ is irrelevant, since one integrates over the $x$-period and over the whole time axis. Then the only $Y$ dependence comes from the terms $\dot x_m(Y),\dot t_m(Y),\ddot x_m(Y),\ddot t_m(Y)$ in   
\beq
\ba{l}
{\cal A}_{YY}(x',t',\phi)=-\ddot x_m(Y){\cal A}_x(x',t',\phi) -\ddot t_m(Y){\cal A}_t(x',t',\phi) \\+2 \dot x_m(Y)\dot t_m(Y){\cal A}_{xt}(x',t',\phi)+{\dot x_m}^2(Y){\cal A}_{xx}(x',t',\phi)+{\dot t_m}^2(Y){\cal A}_{tt}(x',t',\phi), \\
x'=x-x_m(Y), \ \ t'=t-t_m(Y),
\ea
\eeq
implying that
\beq\label{Jm_bis}
\ba{l}
J_m(\lambda_1,Y)=\ddot x_m(Y)\check T[{\cal A}_x]+\ddot t_m(Y)\check T[{\cal A}_t]-2 \dot x_m(Y)\dot t_m(Y)\check T[{\cal A}_{xt}]\\
-{\dot x_m}^2(Y)\check T[{\cal A}_{xx}]-{\dot t_m}^2(Y)\check T[{\cal A}_{tt}],
\ea
\eeq
where
\beq\label{def_Tcheck}
\ba{l}
\check T[f]:=\textcolor{blue}{-}\int\limits_{-\infty}^{\infty}\!\! dt\! \int\limits_0^{L_x}\!\! dx \Big[{\tilde T}_{21} (\lambda_1,x+L_x, x,t) f(x,t)-{\tilde T}_{12} (\lambda_1,x+L_x, x,t) f^*(x,t)\Big].
\ea
\eeq

We illustrate the procedure on the first integral $\check T[{\cal A}_x]$ in \eqref{Jm_bis}. The real part of the integrand is odd with respect to both variables:
\beq
\ba{l}
-4\pi k_x\frac{\sin(k_x x)\,\sin^3(\phi)\left[\sin(\phi)- \cos(k_x x)\cosh(\sigma\!_x t)\right]\sinh(\sigma\!_x t)}{\left[\cosh(\sigma\!_x t) - \sin(\phi)\cos(k_x x)\right]^4},
\ea
\eeq
while the imaginary part is even with respect to both variables: 
\beq
\ba{l}
2 \pi k_x \frac{\left[\cosh(2\sigma\!_x t)+\cos(2\phi)\right]\sin^2(k_x x) \sin^3(\phi)}{\left[\cosh(\sigma\!_x t) - \sin(\phi)\cos(k_x x)\right]^4};
\ea
\eeq
therefore the imaginary part is the only one to give a non-zero contribution. Its $x$-integral is evaluated using standard contour integration techniques, and the subsequent $t$-integration becomes elementary:\\
 \begin{equation}
 	\begin{split}
 	\check T[{\cal A}_x]=\textcolor{blue}{-}&2 \pi i k_x\int\limits_{-\infty}^{\infty}\!\!\! dt \int\limits_0^{L_x}dx \frac{\left[\cosh(2\sigma\!_x t)+\cos(2\phi)\right]\sin^2\!(k_x x) \sin^3\!\phi}{\left[\cosh(\sigma\!_x t) - \sin(\phi)\cos(k_x x)\right]^4}=\\
 		&=4\pi^2 i\int\limits_{-\infty}^{\infty}\!\!\! dt\frac{\sin^3\!\phi\cosh(\sigma\!_x t)}{\left(\sinh^2(\sigma\!_x t) +\cos^2\phi\right)^\frac{3}{2}} =\frac{8i\pi^2\sin^3\phi}{\sigma\!_x}\int\limits_{0}^{\infty}\!\!\! \frac{d\xi}{\left(\xi^2 +\cos^2\phi\right)^\frac{3}{2}}=\\
 		&=4i\pi^2\frac{\sigma\!_x^2}{k_x^5}.
 	\end{split}
 \end{equation}
 For the second integral $\check T[{\cal A}_t]$ in \eqref{Jm_bis}, after remarking that
 \beq
{\cal A}_t=i\left(-2 {\cal A}+{\cal A}_{xx}+2|{\cal A} |^2{\cal A}\right),
\eeq
one proceeds as for the first integral, observing that now it is the real part of the integrand that provides the only non-zero contribution, leading to:\\
\begin{equation}
  \check T[{\cal A}_t]=\frac{20}{3}\pi^2\frac{\sigma\!_x^3}{k_x^6}.
 \end{equation}	
Similarly, one shows that the remaining integrals in \eqref{Jm_bis} are all zero for parity reasons, leading to \eqref{Jm_final}.

\subsection{The two-step recurrence of Q1D AWs}

Introducing the slowly varying analogue $\{Q_m(Y)\}_{m\in\ZZ}$ of the sequence \eqref{def_Qm} of normalized squared gaps:
\beq\label{def_Qm}
\ba{l}
Q_m(Y)=\frac{\lambda_1^2}{\eps^2}\mbox{Gap}^2_m(Y), \\
Q_0(Y)=\alpha_0(Y)\alpha_1(Y),
\ea
\eeq
one obtains from \eqref{variation_trT_gap} and \eqref{Jm_final} the following recurrence of $O(1)$ slowly varying functions:
\beq\label{variation_Qm(Y)}
\ba{l}
Q_{m}(Y)=Q_{m-1}(Y)+b \left(\frac{\delta}{\eps}\right)^2 8\cos\phi \sin^2\phi \left[\frac{10}{3}\sin\phi \,\ddot t_m(Y)+i\, \ddot x_m(Y)\right]\\
=Q_{m-1}(Y)+b \left(\frac{\delta}{\eps}\right)^2\! \left(\frac{\sigma\!_x}{k_x}\right)^2\!\!\left[\frac{5}{3}\sigma_x \,\ddot t_m(Y)+i\, k_x\,\ddot x_m(Y)\right], \ m\ge 1, \ \textcolor{red}{b}=\pm 1,
\ea
\eeq
either written in terms of the angle $\phi=\arccos(k_x/2)$ or, maybe better, in terms of the unstable wave number $k_x$ and growth rate $\sigma\!_x$ of the unperturbed NLS equation.

The other important ingredient is the slowly varying version of the formula \eqref{variation_xm_tm} connecting the two subsequent nonlinear stages of MI described by $u_{m}$ and $u_{m+1}$ and the gap $\mbox{Gap}_m(Y)$ (or $Q_m(Y)$, via \eqref{def_Qm}), essentially constant during the logarithmically large time interval connecting these nonlinear stages:
\beq\label{variation_xm(Y)_tm(Y)}
\ba{l}
t_{m+1}(Y)-t_{m}(Y)=\frac{1}{\sigma\!_x}\ln\!\left(\frac{\sigma{\!_x}\!^4}{4\eps^2 |Q_m(Y) |} \right), \\
x_{m+1}(Y)-x_{m}(Y)=\frac{\arg(Q_m(Y))}{k_x}, \ \ \mbox{mod }L_x .
\ea
\eeq

These equations hold true for $m=0$, as it can be checked from  \eqref{LSMIa}, \eqref{u_m(Y)}, \eqref{def_Qm}, and, due to the space/time translation invariance of the MNLS models under investigation, they generalizes to arbitrary $m\in\ZZ$, or, to be more precise, for a number of recurrences before the perturbative term $\gamma\delta^2 u_{YY}$ becomes relevant, modifying the picture.
\vskip 10pt
Summarizing, the $m^{th}, \ m\ge 1$, NLSMI, described at leading order by the adiabatic deformation
\beq\label{def_um(Y)}
u_m(x,Y,t)={\cal A}\left(x-x_m(Y),t-t_m(Y),\phi\right)e^{2it+2i\phi_m}, \ \phi_m=(2m-1)\phi 
\eeq
of the quasi-homoclinic AB \eqref{Akhmed}, is characterized by the slowly varying functions $\{x_m(Y),t_m(Y)\}$, while the LSMI preceeding it is characterized  by the slowly varying gap ${Gap}_{m-1}(Y)$ (or by $Q_{m-1}(Y)$). Then equations \eqref{variation_Qm(Y)},\eqref{variation_xm(Y)_tm(Y)}, together with the starting points
\beq\label{initial}
\ba{l}
Q_0(Y)=\alpha_0(Y)\alpha_1(Y), \\
t_1(Y)=\frac{1}{\sigma\!_x}\ln\!\left(\frac{\sigma\!_x^2}{2\eps |\alpha_1(Y)|} \right), \ \
x_1(Y)=\frac{\arg(\alpha_1(Y))}{k_x}+\frac{L_x}{4}, \ \mbox{mod }L_x, 
\ea
\eeq
describe, at leading order, the two-step recurrence of $x$-periodic Q1D AWs of the ENLS and HNLS equations in $2+1$ dimensions as follows. To go from $\{Q_{m-1}(Y),x_m(Y),t_m(Y)\}$ to $\{Q_{m}(Y),x_{m+1}(Y),t_{m+1}(Y)\}$, one first uses \eqref{variation_Qm(Y)} to construct $Q_m(Y)$ (the continuous arrows of Figure \ref{recurrence}), and then \eqref{variation_xm(Y)_tm(Y)} to construct $\{x_{m+1}(Y),t_{m+1}(Y)\}$ (the dashed arrows of Figure \ref{recurrence}).
\begin{figure}[h!!!!!!!!!!!!!!!!!!!!!!!!!!!!!!]
  \includegraphics[width=13cm,height=4cm]{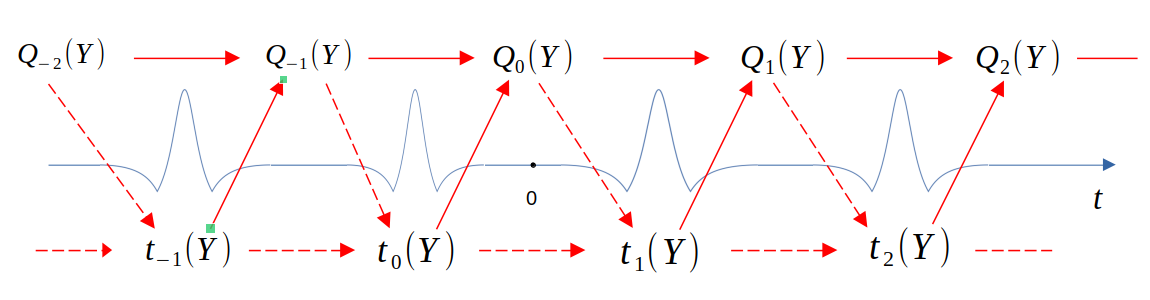}
  \caption{The two-step recurrence scheme describing the AW recurrence of Q1D AWs of the ENLS and HNLS equations. On the time axis we see the schematic recurrence of linear and nonlinear stages of MI.}\label{recurrence}
\end{figure}

We remark that the recurrence \eqref{variation_Qm(Y)},\eqref{variation_xm(Y)_tm(Y)} can be reformulated as follows. Let
\beq\label{def_Em_Qm}
\ba{l}
E_m(Y):=e^{-\sigma\!_x t_m(Y)+ik_x x_m(Y)}, \ \ m\ge 0, \\
\tilde Q_m(Y):=-\left(\frac{\mbox{Gap}_m \!(Y)}{k_x \sigma\!_x} \right)^2, 
\ea
\eeq
then the two real equations \eqref{variation_xm(Y)_tm(Y)} become the single complex equation
\beq
E_{m+1}(Y)=\tilde Q_m(Y)E_{m}(Y),
\eeq
and \eqref{variation_Qm(Y)} is rewritten as
\beq
\tilde Q_{m}(Y)=\tilde Q_{m-1}(Y)+4\,b \left(\frac{\delta}{k_x\sigma\!_x} \right)^2\left[\frac{5}{3}\sigma\!_x\, \ddot t_m(Y)+i k_x\, \ddot x_m(Y) \right].
\eeq

Equations \eqref{variation_Qm(Y)},\eqref{variation_xm(Y)_tm(Y)} show four basic novel properties of the AW recurrence in the Q1D regime.\\
i) While the first NLSMI is essentially the same for all the MNLS equations, the second and the subsequent nonlinear stages of MI are different for the different MNLS equations ($\gamma=\pm 1$ for the ENLS and HNLS equations in \eqref{variation_Qm(Y)}). \\
ii) The two independent small parameters $\eps$ and $\delta$ appear through the ratio $(\delta/\eps)^2$ in \eqref{variation_Qm(Y)}. It means that, no matter how small is the parameter $\delta$ of the quasi one dimensionality, if $\eps$ is such that $\delta/\eps=O(1)$, then there are $O(|J_m|)$ changes in the dynamics.\\
iii) In general, the slowly varying functions $\{x_m(Y),t_m(Y) \}$ undergo significant changes during the recurrence described by \eqref{variation_Qm(Y)},\eqref{variation_xm(Y)_tm(Y)}; it follows that two subsequent nonlinear stages of MI for the same MNLS equation are qualitatively and quantitatively different, exhibiting different combinations of fission and fusion processes.\\
iv) Together with the condition $\ln(1/\eps)\ll 1/\delta$, necessary to ensure the validity of the above analytic description for a number of recurrences, the second important requirement is self-consistency: the dynamics described by equations \eqref{variation_Qm(Y)}-\eqref{initial} must always be in agreement with the hypotheses under which equations \eqref{variation_Qm(Y)}-\eqref{initial} have been derived. If, for instance, a substantial narrowing in the $Y$ direction occurs during the evolution, then the Q1D regime might cease to be valid, presumably together with the above formulas.

\section{Local description of fission and fusion}\label{local}

As in \cite{CS_Q1D_NLSMI1}, to easily understand the basic features of Q1D AW fission, fusion, and recurrence, we restrict the family of initial data in \eqref{Cauchy_data_Q1D} to
\beq\label{FourierCoeff(Y)_simpler}
c_{\pm}(Y)=c_{0\pm}f(Y)e^{\mp i g(Y)},
\eeq
where $c_{0\pm}$ are arbitrary complex constants, and $f(Y),g(Y)$ are real functions of $Y$, assuming without loss of generality that $0< f(Y)\le 1$. Then
\beq\label{alpha10(Y)_alpha11(Y)_simpler}
\ba{l}
\alpha_1(Y)=\alpha_{10} f(Y)e^{i g(Y)}, \ \ \alpha_0(Y)=\alpha_{00} f(Y)e^{-i g(Y)}, \\
\alpha_{10}=e^{-i\phi}c^*_{0+}-e^{i\phi}c_{0-}, \ \ \alpha_{00}=e^{i\phi}c^*_{0-}-e^{-i\phi}c_{0+},
\ea
\eeq
the starting point \eqref{initial} of the AW recurrence is given by
\beq\label{init(Y)_rec_simpler}
\ba{l}
Q_0(Y)=\alpha_{00}\alpha_{10}f^2(Y), \\
t_1(Y)=t_{10}+\frac{1}{\sigma_x}\ln\!\left(\frac{1}{f(Y)} \right), \ \ t_{10}=\frac{1}{\sigma_x}\ln\!\left(\frac{\sigma_x^2}{2\eps |\alpha_{10} |} \right), \\
x_1(Y)=x_{10}+\frac{g(Y)}{k_x}, \ \ x_{10}=\frac{\arg\alpha_{10}}{k_x}+\frac{L_x}{4}, \ \mbox{mod }L_x,
\ea
\eeq
and the first NLSMI is described at leading order by
\beq\label{u1_simpler}
u_1(x,Y,t)=e^{2it+2i\phi}{\cal A}\left(x-x_1(Y),t-t_1(Y),\phi\right).
\eeq

The extremal points of $f(Y)$ play the main role in the analysis. Suppose that $Y_0$ is a local max of $f(Y)$:
\beq\label{about_YM}
\ba{l}
f(Y)=f(Y_0)-\frac{|f''(Y_0)|}{2}(Y-Y_0)^2 +O((Y-Y_0)^3), \ \ \ |Y-Y_0|\ll 1, \\
g(Y)=g(Y_0)+g'(Y_0)(Y-Y_0)+\frac{g''(Y_0)}{2}(Y-Y_0)^2+O((Y-Y_0)^3).
\ea
\eeq
Then $t_1(Y)$ has a local minimum at $Y=Y_0$:
\beq
\ba{l}
t_1(Y)\sim t_{0}+\frac{|f''(Y_0)|}{2\sigma\!_x f(Y_0)}(Y-Y_0)^2, \ \ |Y-Y_0|\ll 1, \\
t_{0}=\frac{1}{\sigma_x}\ln\!\left(\frac{\sigma\!_x^2}{2\eps |\alpha_{10}|f(Y_0)} \right),
\ea
\eeq
and
\beq
\ba{l}
x_1(Y)\sim x_0 +\frac{g'(Y_0)}{k_x}(Y-Y_0)+\frac{g''(Y_0)}{2 k_x}(Y-Y_0)^2, \ \ |Y-Y_0|\ll 1, \\
x_0 = x_{10}+\frac{g(Y_0)}{k_x}.
\ea
\eeq
Since the AW reaches its amplitude maximum along the curves $x=x_1(Y)$ and $t=t_1(Y)$ during the first NLSMI described by $u_1(x,Y,t)$, at $t=t_{0}$ in $(x,Y)=(x_0,Y_0)$ the AW splits locally into two waves of amplitude $1+2\sin\phi$, traveling on the trajectories  $x=x_1(Y^{\pm}(t))$, where $Y^{\pm}(t)\sim Y_0\pm\sqrt{\frac{2\sigma\!_x f(Y_0)}{|f''(Y_0)|}(t-t_{0})}, \ 0\le t-t_{0}\ll 1$. Therefore the fission time is $t_{fiss}=t_{0}$, and the two fission products separate with infinite speed at fission time according to the universal law (see Figure \ref{schema_fiss}):
\beq\label{fission_speed_2+1}
\dot{Y}^{\pm}(t)=\pm \sqrt{\frac{\sigma\!_x f(Y_0)}{2|f''(Y_0)| }}\frac{1}{\sqrt{t-t_{fiss}}}, \ \ 0\le t-t_{fiss}\ll 1.
\eeq
\begin{figure}
\centering
\includegraphics[width=11.0cm,height=13.5cm]{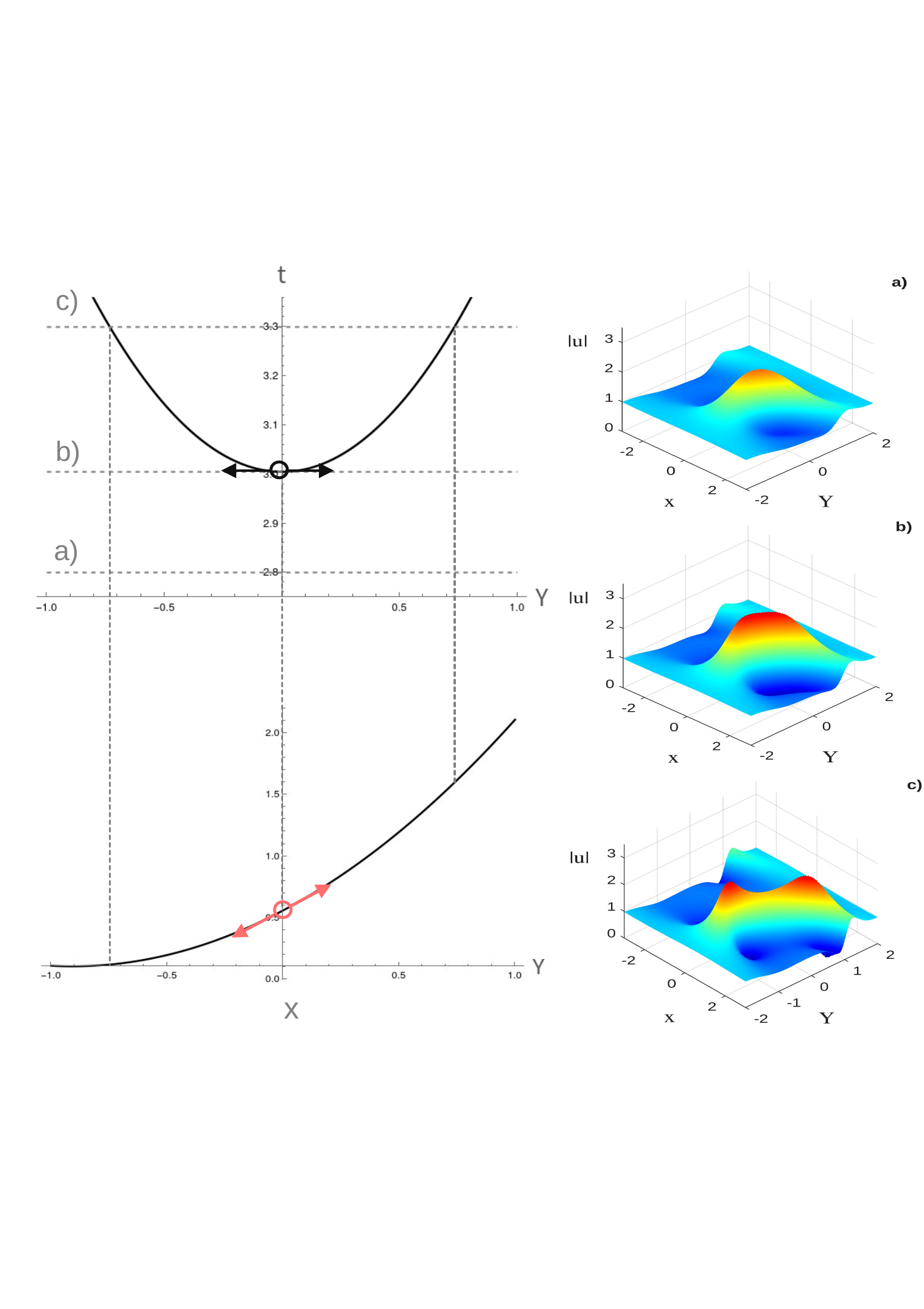}
\caption{A local description of the process of ``growth from the background + fission'' for $f(Y)=1-0.8 Y^2+O(Y^3)$, $g(Y)=Y+0.5 Y^2+O(Y^3)$, $|Y|\ll 1$; the upper left figure shows the graph of $t_1(Y)$ in the $(Y,t)$ plane, together with the horizontal dashed lines $t=2.80,t_{fiss}=3.005,3.30$, and the lower left shows the trajectory $x_1(Y)$ of the fission products in the $(Y,x)$ plane. The right figures show the three snapshots of $|u_1(x,Y,t)|$ at the times indicated by the dashed horizontal lines of the upper left figure. From up to down: $t=2.80$: the AW growth from the background in $(x,Y)=(x_{0},0)$, $x_{0}=0.559$; $t=t_{fiss}=3.005$: the AW fission in $(x,Y)=(x_{0},0)$; $t=3.20$; the separated fission products in $(x_1(Y^+),Y^+)=(0.591,0.735)$ and $(x_1(Y^-),Y^-)=(0.121,-0.735)$. Here $\eps=0.01, \ L_x=6, \ \alpha_{10}=0.421-0.622 i .$}\label{schema_fiss}
\end{figure}
Viceversa, if $Y_0$ is a local min of $f(Y)$, then $t_1(Y)$ has a local max at $Y=Y_0$: $t_1(Y)\sim t_{0}-\frac{|f''(Y_0)|}{2\sigma\!_x f(Y_0)}(Y-Y_0)^2$, $|Y-Y_0|\ll 1$, and a local fusion takes place at $t=t_{fus}=t_{0}$ in $(x_1(Y_0),Y_0)$: two waves travel against each other with trajectories  $x=x_1(Y^{\pm}(t))$, $Y^{\pm}(t)\sim Y_0\mp\sqrt{\frac{2\sigma\!_x f(Y_0)}{|f''(Y_0)|}(t_{fus}-t)}$, $0\le t_{fus}-t\ll 1$, and undergo fusion  with infinite speed at fusion time according to the universal law:
\beq\label{fission_speed_2+1}
\dot{Y}^{\pm}(t)=\mp \sqrt{\frac{\sigma\!_x}{2|f''(Y_0)| }}\frac{1}{\sqrt{t_{fus}-t}}, \ \ 0\le t_{fus}-t\ll 1,
\eeq
followed by a decay to the background (see Figure \ref{schema_fus}). Therefore also fusion is a critical process.
\begin{figure}
	\centering
	\includegraphics[width=11.0cm,height=13.5cm]{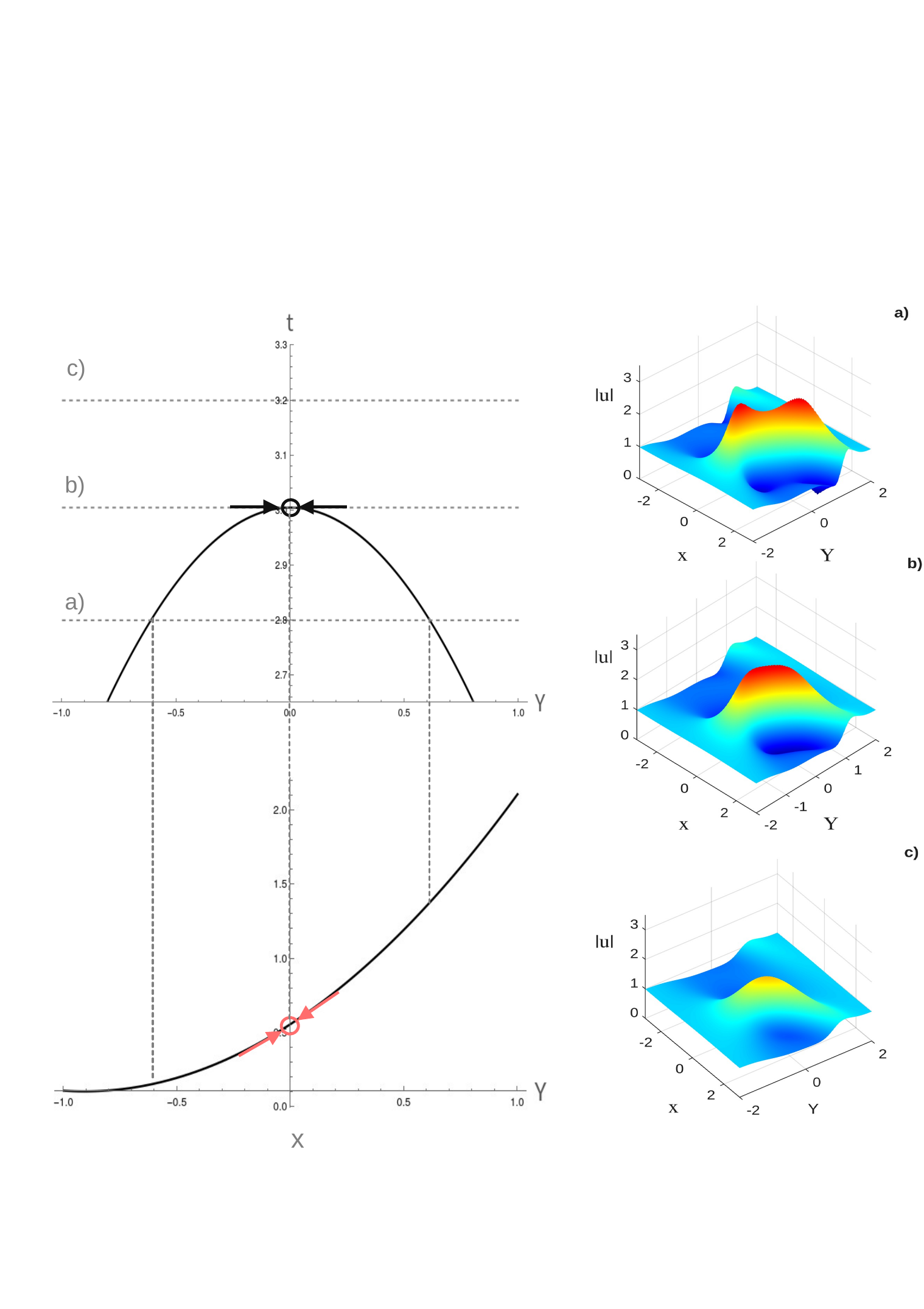}
	\caption{A local description of the process of ``fusion + decay to the background'' for $f(Y)=1+0.8 Y^2+O(Y^3)$, $g(Y)=Y+0.5 Y^2+O(Y^3)$, $|Y|\ll 1$; the upper left figure shows the graph of $t_1(Y)$, together with the horizontal dashed lines $t=2.80,3.005,3.20$, and the lower left shows the trajectory $x_1(Y)$ of the two waves undergoing fusion in the $(Y,x)$ plane. The right figures show the three snapshots of $|u_1(x,Y,t)|$ at the times indicated by the dashed horizontal lines of the upper left figure. From up to down:  $t=2.980$: the two waves centered at  $(x(Y_+),Y_+)=(1.374,0.610)$ and at $(x(Y_-),Y_-)=(0.154,-0.610)$ travel against each other; $t=t_{fus}=3.005$: the fusion in $(x,Y)=(x_{0}=0.559,0)$; $t=3.030$: the AW decay to the backgroung, centered at $(x,Y)=(x_{0},0)$, $x_{0}=0.559$. Here $\eps=0.01, \ L_x=6, \ \alpha_{10}=0.421-0.622 i .$}\label{schema_fus}
\end{figure}

We conclude that \\
\textit{the two processes of ``AW growth from the background + fission'' and ``fusion + AW decay to the background'' are the basic ingredients of the first NLSMI, and of the subsequent ones, corresponding to the local minima and maxima of $t_m(Y)$, while the curves $x=x_m(Y)$ are the trajectories of the fission products. Varying the initial data, functions $t_m(Y)$ and $x_m(Y)$ vary according to equations \eqref{LSMIb}, \eqref{u_m(Y)},  giving rise to various combinations of these two basic processes, and then to rich and often aesthetically pleasing choreographies}.

The above considerations imply that, in the Q1D regime, a growing AW undergoes fission when it reaches its maximum elevation, although this synchronization is not expected a priori, since the two events are defined in two different ways. At fission, the $y$-curvature of the elevation changes sign according to the law
\beq\label{def_tfiss}
\mbox{sign}\!\left(|u|_{yy}(0,0,\tau)\right)=\mbox{sign}\!\left(\tau\right) \ \ (\mbox{with \ sign}(0)=0), \ \tau=t-t_{fiss},
\eeq
while the maximum elevation is reached at time $t_{max}$ for which $|u|_t(0,0,t_{max})=0$, and $|u|$ is not necessarily the same at $t_{max}$ and $t_{fiss}$. To better understand this synchronization, we Taylor expand \eqref{init(Y)_rec_simpler} and \eqref{u1_simpler} about the fission space/time point $(x_{fiss},Y_{fiss},t_{fiss})=(x_0,Y_0,t_0)$, obtaining the following local description of the fission process (choosing $\dot x_1(Y_0)=0$ for parity reasons):
\beq\label{local_fission_Q1D}
\ba{l}
|u_1(x,Y,t)|^2\sim \rho_{00}- \rho_{02}\, (\tau-\frac{\ddot t_1(Y_{fiss})}{2}\eta^2)^2-\rho_{20}\, (\xi-\frac{\ddot x_1(Y_{fiss})}{2}\eta^2)^2\\
\sim \rho_{00}- \rho_{02}\tau^2+\rho_{02}\ddot t_1(Y_{fiss})\tau \eta^2-\left(\rho_{20}\xi^2+\frac{\rho_{02}\ddot t_1^2(Y_{fiss})}{4}\eta^4 \right),\\
\xi=x-x_{fiss}, \ \ \eta=Y-Y_{fiss}, \ \ \tau=t-t_{fiss},
\ea
\eeq
where
\beq\label{coeff_local_fission_Q1D}
\ba{l}
\rho_{00}=a_0^2, \ \ \rho_{02}=4 a_2 \sin^2\phi, \ \ \rho_{20}=a_0\, a_2,\\
a_0=1+2\sin\phi, \ \ a_2=\frac{8\sin\phi\cos^4\!\phi}{(1-\sin\phi)^2}
\ea
\eeq
(see Figure \ref{wave-fission-3D}).
\begin{figure}[h!!!!]
  \centering
  \includegraphics[width=4cm]{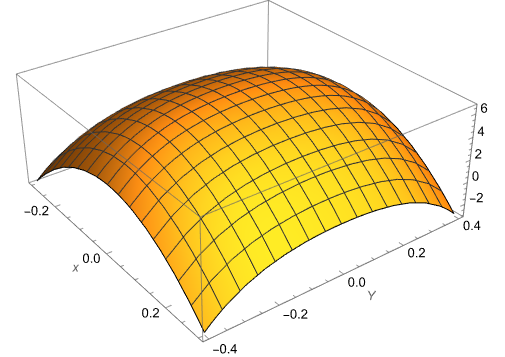} \ \includegraphics[width=4cm]{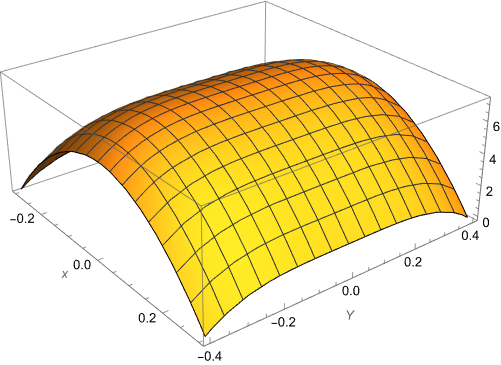} \    
  \includegraphics[width=4cm]{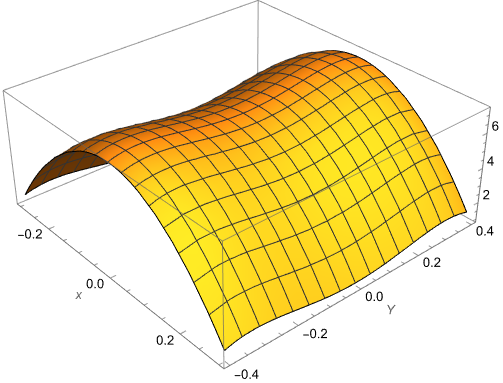}
	\caption{Three snapshots of  $|u_1(x,Y,t)|^2$ describing locally the fission process in the $Y$ direction. From left to right, $t=-0.1$: the AW growth, $t=0$ the AW fission, $t=0.1$: the separation of the fission products. Here $L_x=0 \Rightarrow \phi=1.02$, $\ddot{t}_1(0)=2$, $\dot{x}_1(0)=\ddot{x}_1(0)=0$.}\label{wave-fission-3D}
\end{figure}

The coefficient $\rho_{00}$ is the square of the wave amplitude at fission, the term $-\rho_{02}\, \tau^2$ describes the growth and decay of the AW before and after fission, the term $\rho_{02}\,\ddot t_1(Y_{fiss})\,\tau\, \eta^2$ is responsable for the $y$-curvature change of sign in the fission space-time point, and the term $-\left(\rho_{20}\,\xi^2+\frac{\rho_{02}\,\ddot t_1^2(Y_{fiss})}{4}\eta^4 \right)$ describes the overall space localization of the wave (see Figure \ref{wave-fission-3D}).

From \eqref{local_fission_Q1D} it is easy to verify that $|u|_{yy}(0,0,\tau)\sim \rho_{02}\,\ddot t_1(Y_{fiss})\,\tau$, in agreement with \eqref{def_tfiss}, and that $|u|_t(0,0,\tau)\sim -\rho_{02}\tau$, implying that the fission and maximum amplitude events coincide, for the AW amplitude $\sqrt{\rho_{00}}=1+2\sin\phi$.

Changing $\tau\to -\tau$ and fiss $\to$ fus, equations  \eqref{local_fission_Q1D},\eqref{coeff_local_fission_Q1D} give the local description of fusion.

Since AW fission is not restricted to the Q1D regime \cite{CS_Q1D_NLSMI1}, equation \eqref{local_fission_Q1D} with more general coefficients: 
\beq\label{local_fission_gen}
|u(x,y,t)|^2\sim c_0-c_1\,\tau-c_2 \tau^2+ c_3\,\tau\,\eta^2-(c_4 \xi^2+c_5 \eta^4),
\eeq
where $c_j>0, \ j=1,\dots 5$, provides the universal and local description of it \cite{CS_2+1_phen}, and we see in \eqref{local_fission_gen} the presence of the term $-c_1 \tau$, absent in the Q1D regime, indicating that the maximum amplitude event takes place earlier, at time $\tau=-c_1/(2 c_2)<0$, with amplitude $|u|\sim\sqrt{c_0+c_1^2/(4 c_2)}$ greater than the amplitude $|u|\sim\sqrt{c_0}$ at fission time $\tau=0$ \cite{CS_Q1D_NLSMI1}.

\section{Illustrative examples}\label{Examples}

The recurrence of the nonlinear stages of MI for $x$-periodic Q1D AWs, described by equations \eqref{def_um(Y)},\eqref{variation_Qm(Y)},\eqref{variation_xm(Y)_tm(Y)}, and \eqref{initial}, is here illustrated on two basic examples: i) a doubly periodic Q1D AW, and ii) an $x$-periodic Q1D AW localized over the background in the $y$ direction, restricting again our considerations to formulas \eqref{FourierCoeff(Y)_simpler}-\eqref{u1_simpler}, in the simplest case of a single global max (in $Y=0$) of $f(Y)$. These two examples are two of the cases considered in our previous work \cite{CS_Q1D_NLSMI1} in which we concentrated on the first NLSMI, viewed in this paper as just the first of a sequence of nonlinear stages of MI.

\subsection{The recurrence of a  doubly periodic Q1D AW}\label{doubly_periodic}

Consider an even and periodic function $f(Y)$, with a single hump inside the period $L_Y$ \cite{CS_Q1D_NLSMI1}:
\beq\label{example}
f(Y)=\frac{1+\beta\cos(k_Y Y)}{1+\beta}, \ \ g(Y)=0,  \ \ k_Y=\frac{2\pi}{L_Y}, \ \ 0<\beta<1.
\eeq

As discussed in \cite{CS_Q1D_NLSMI1}, \eqref{init(Y)_rec_simpler} implies that $t_1(Y)$ possesses minima in $Y_n=n L_Y$, with $t_1(Y_n)=t_{10}$, and maxima in $\tilde Y_n=(n+1/2)L_Y$, with $t_1(\tilde Y_n)=t_{10}+\ln\!\left(\frac{1+\beta}{1-\beta} \right)$, $n\in\ZZ$ (see Figure \ref{NLSMI_t1}), while $x_1(Y)=x_{10}$. Consequently growing AWs undergo fission at $t_{fiss}=t_{10}$ in $(x_{10},Y_n)$, $n\in\ZZ$; then the right fission product in $Y_n$ and the left fission product in $Y_{n+1}$ travel against each other along the straight line $x=x_{10}$, and undergo fusion in $(x_{10},\tilde Y_n)$ at $t_{fus}=t_1(\tilde Y_n)$ into an AW decaying to the background (see Figure \ref{NLSMI1}). As we know, this first NLSMI is the same for the elliptic and hyperbolic NLS equations, and for all MNLS equations.

\begin{figure}[h!!!!!!!!!!!!!!!!!!!!!!!!!!!!!!]
		\centering
		\includegraphics[width=9cm,height=6cm]{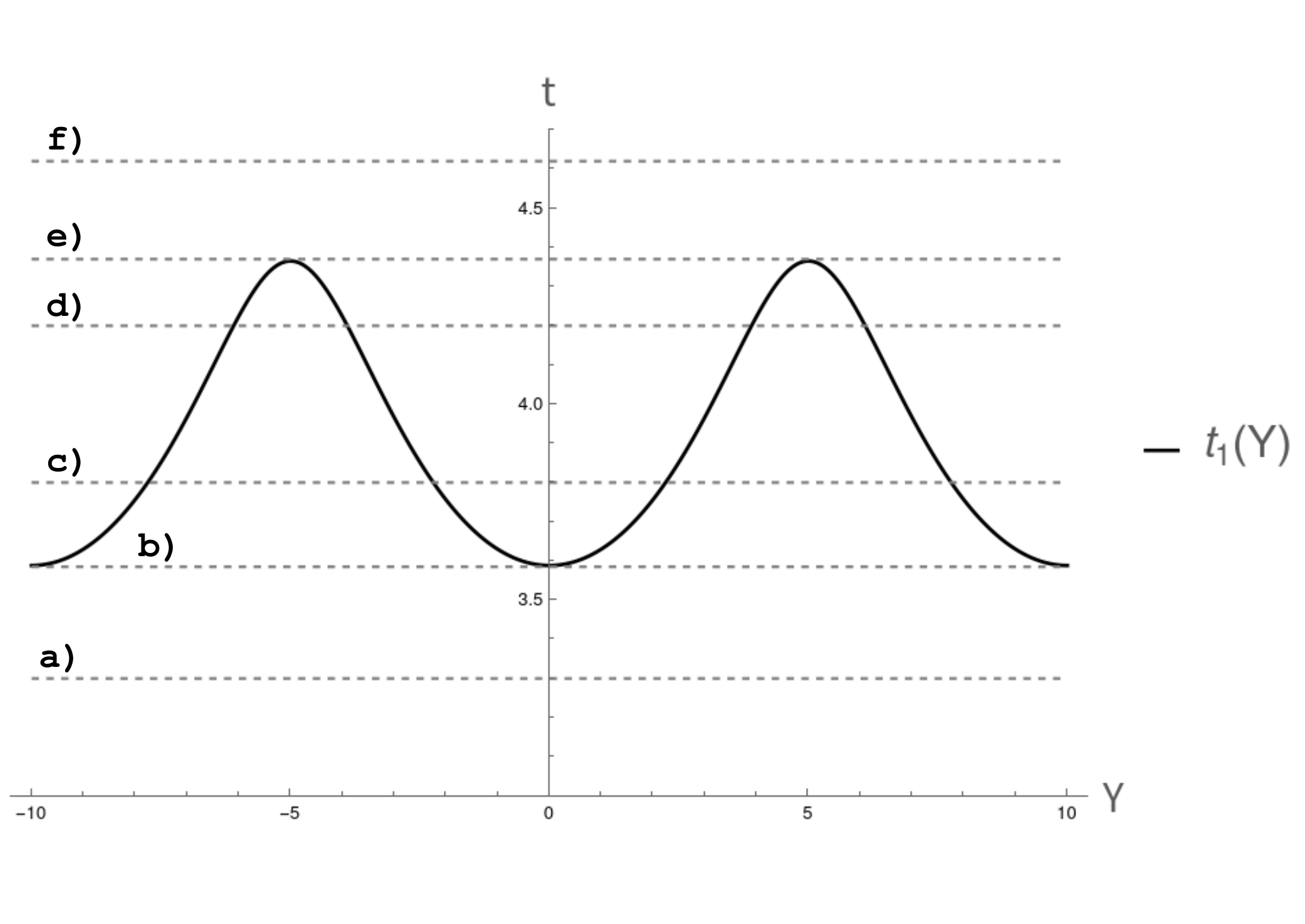}
		\caption{The first NLSMI for both HNLS and ENLS equations. The graph of $t_1(Y)$ in the $(Y,t)$ plane, together with the six horizontal dashed lines $t=3.30$, $3.59$, $3.80$, $4.20$, $4.37$, $4.62$ corresponding to the snapshots of Figure \ref{NLSMI1}. For the sake of clarity, the period in the Y-direction is doubled: $-L_Y\le Y\le L_Y$, with $L_x=6, \ L_Y=10$, and  $c_{0-}=0.3+i 0.1$, $c_{0+}=0.084$, $\beta=0.3$, $\eps=10^{-2}, \ \delta=10^{-3}$.}\label{NLSMI_t1}
	\end{figure} 
\begin{figure}[h!!!!!!!!!!!!!!!!!!!!!!!!!!!!!!]
		\centering
		\includegraphics[width=12cm,height=7cm]{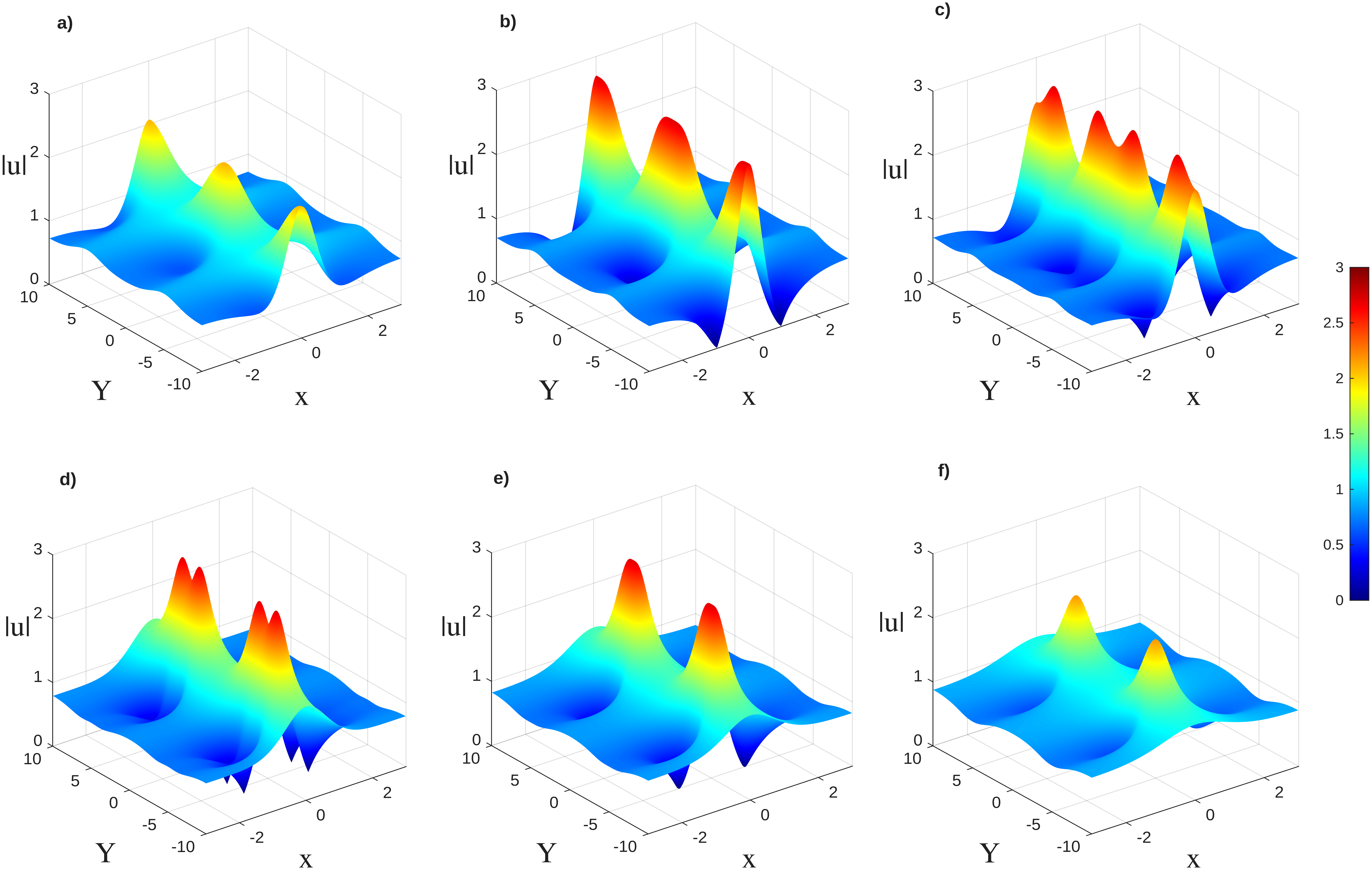}
		\caption{The first NLSMI for both HNLS and ENLS equations. Six snapshots  of $|u(x,Y,t)|$ taken at times $t=3.30$, $3.59$, $3.80$, $4.20$, $4.37$, $4.62$, over the rectangle $-L_x/2\le x\le L_x/2$ and $-L_y\le Y\le L_y$, with the same data as in Figure \ref{NLSMI_t1}. From left to right: a) $t=3.30$: the AW growth; b) $t=3.59$: its fission time; c) $t=3.80$: initial separation of the fission products; d) $t=4.20$ the left fission product of the AW at $Y=0$ and the right fission product of the AW at $Y=L_Y$ get closer; e) $t=4.37$: their fusion; f) $t=4.62$: the AW decay to the background.}\label{NLSMI1}
	\end{figure}  
 \newpage       
        The second NLSMI is richer than the first one, and different for the ENLS and HNLS equations, and it is illustrated in Figures \ref{NLSMI2_HNLS_t2_x2}, \ref{NLSMI2_HNLS}, and \ref{NLSMI2_comparison_HNLS_ENLS}, with the same initial data used to describe the first NLSMI in Figures \ref{NLSMI_t1} and \ref{NLSMI1}. While the trajectory of the fission products during the first NLSMI is constrained to the straight line $x=x_{10}$, since $x_1(Y)=x_{10}$, now their trajectory, described by equation $x=x_2(Y)$, is shown in Figure \ref{NLSMI2_HNLS_t2_x2}. In addition, $t_2(Y)$ possesses global minima in $Y_n=n L_Y$, like $t_1(Y)$, but the global maxima of $t_1(Y)$ in $\tilde Y_n=(n+1/2)L_Y$ are local minima for $t_2(Y)$, since new global maxima arise in between (see Figure \ref{NLSMI2_HNLS_t2_x2}).

\begin{figure}
	\centering
	\includegraphics[width=13cm,height=15cm]{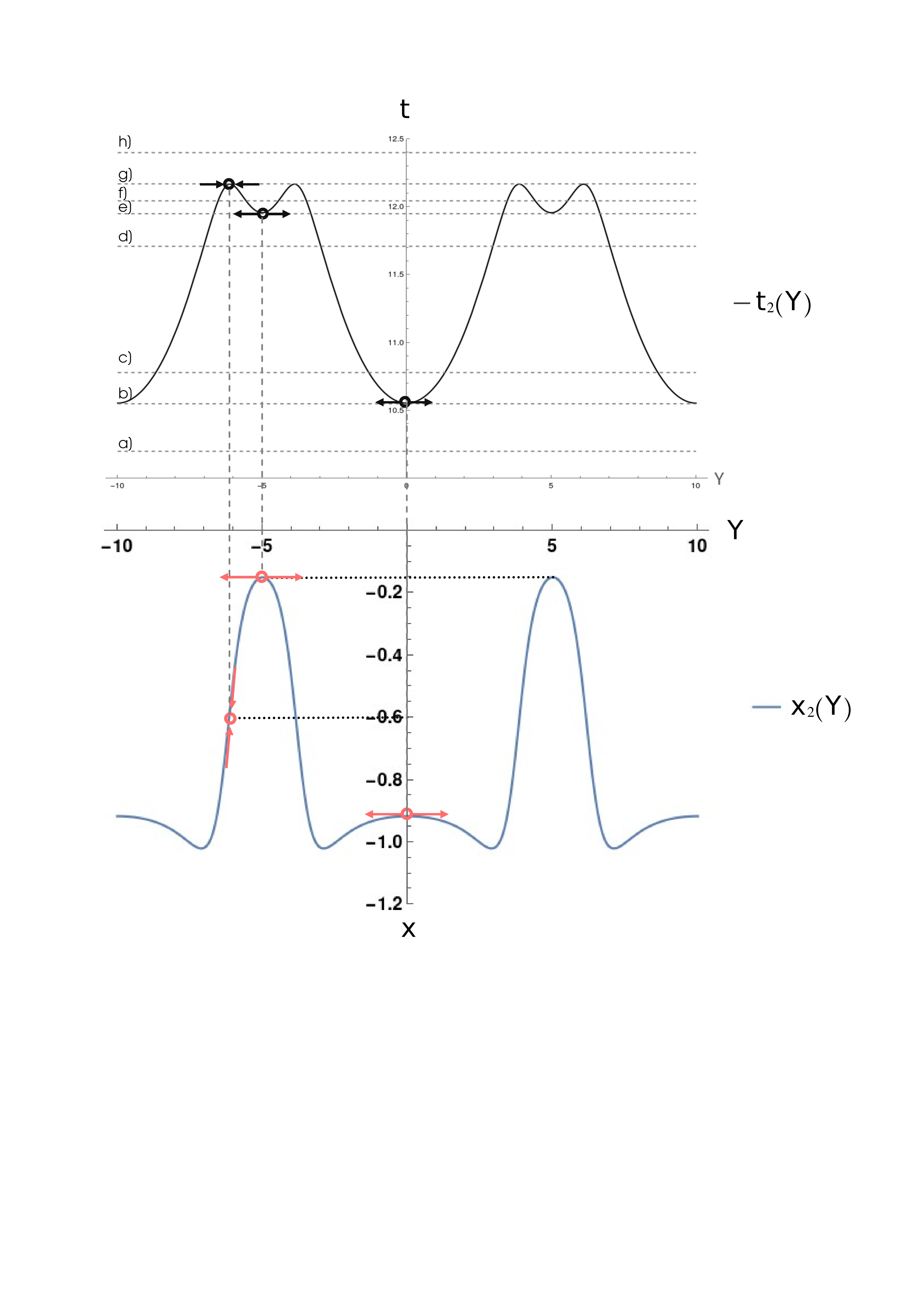}
	\caption{The second NLSMI for the HNLS equation. The upper picture shows the graph of $t_2(Y)$ in the $(Y,t)$ plane, together with the height horizontal dashed lines $t=10.2,\, 10.55,\, 10.78,\, 11.71,\, 11.95,\, 12.05,\, 12.17,\, 12.4$ corresponding to the snapshots of Figure \ref{NLSMI2_HNLS}. The lower picture  shows the trajectory $x_2(Y)$ of the fission products in the $(Y,x)$ plane. Outward-pointing and inward-pointing arrows indicate respectively fission and fusion events. For the sake of clarity, the period in the Y-direction is doubled: $-L_y\le Y\le L_y$, with $L_x=6, \ L_Y=10$, and with the same data as in Figure \ref{NLSMI_t1}}\label{NLSMI2_HNLS_t2_x2}
      \end{figure}

      \newpage
      
      Consequently growing AWs undergo fission at $t_{fiss}=t_2(0)$ in $(x_{2}(0),Y_n)$; then the internal fission products between $Y_n$ and  $Y_{n+1}$ travel along the curve $x=x_2(Y)$, getting closer. At time $t_2(L_Y/2)$ a new AW reaches it max in $(x_2(L_Y/2),L_Y/2)$, and undergoes fission. At last, the internal fission products of the first AW undergo fusion with the fission products of the second AW. After fusion, the second NLSMI ends with the decay to the background (see Figure \ref{NLSMI2_HNLS}).
      
\begin{figure}[h!!!!!!!!!!!!!!!!!!!!!!!!!!!!!!]
	\centering
	\includegraphics[width=15cm,height=7cm]{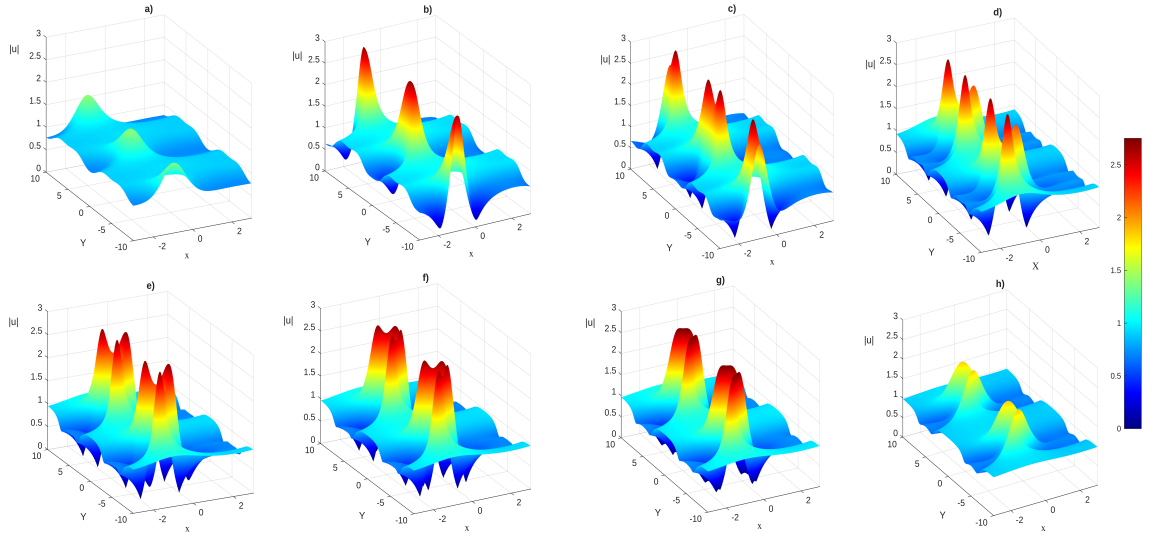}
	\caption{The second NLSMI for the HNLS equation.  Eight snapshots of $|u(x,Y,t)|$ taken at times $t=10.2$, $10.55$, $10.78$, $11.71$, $11.95$, $12.05$, $12.17$, $12.4$, over the rectangle $-L_x/2\le x\le L_x/2$ and $-L_y\le Y\le L_y$, with the same data as in Figure \ref{NLSMI_t1}. a) $t=10.2$: AW growth from the background, b) $t=10.55$: its fission time, c) $t=10.78$: separation of its fission products, d) $t=11.71$: growth of a second AW, e) $t=11.95$: its fission time, f) $t=12.05$: separation of its fission products, g) $t=12.17$: oblique fusion of the fission products of the two AWs, h) $t=12.4$: decay to the background.}\label{NLSMI2_HNLS}
      \end{figure}
      In this experiment, the second NLSMI of the ENLS equation is qualitatively similar to that of the HNLS equation, but with $O(1)$ differences (see Figure \ref{NLSMI2_comparison_HNLS_ENLS}).
   \newpage    
      \begin{figure}[h!!!!!!!!!!!!!!!!!!!!!!!!!!!!!!]
	\centering
	\includegraphics[width=14cm,height=15cm]{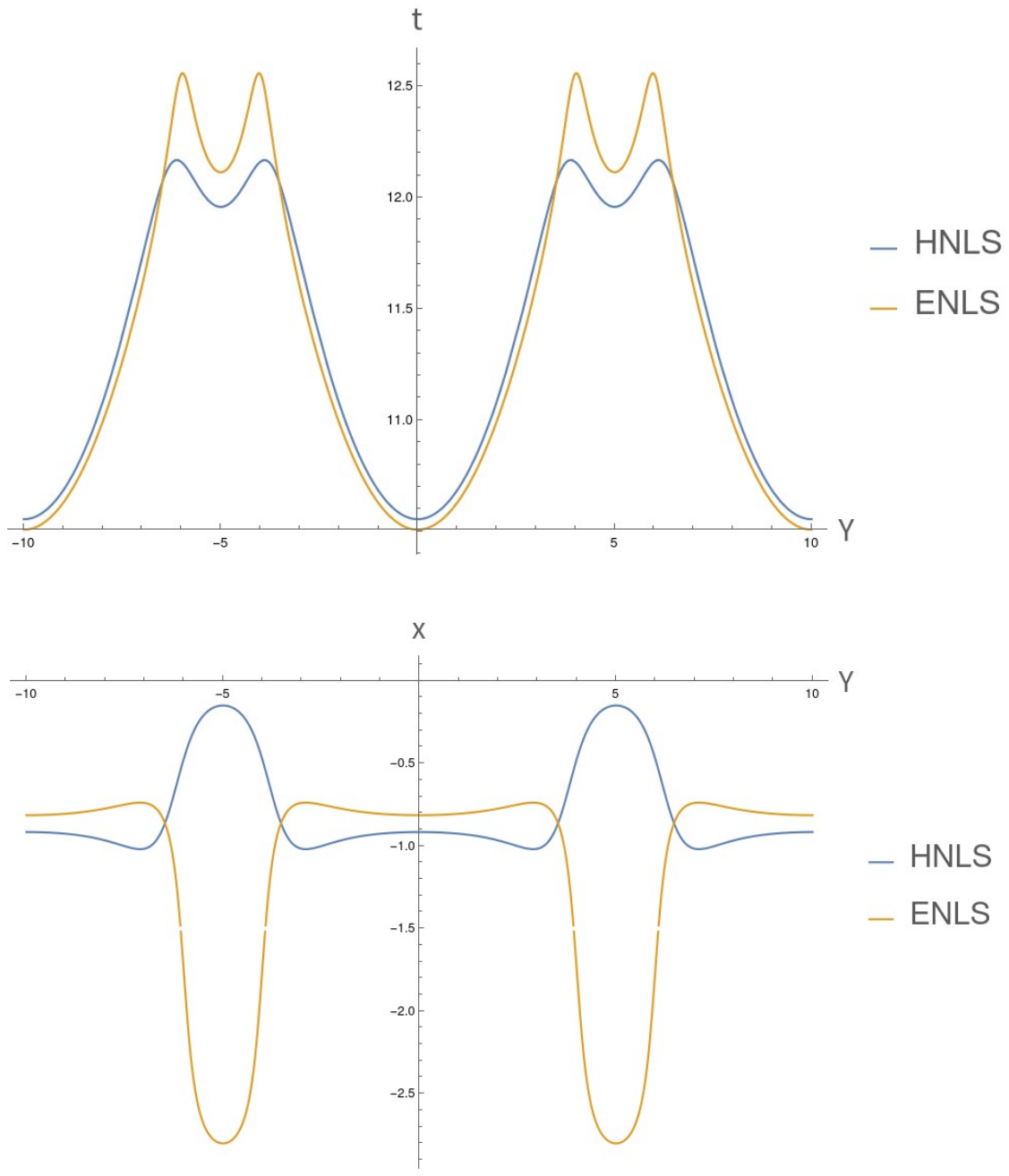}
	\caption{Comparison of the second NLSMI for the ENLS and HNLS equations. The upper figure shows the plots of $t_2(Y)$ for both equations; the lower figure shows the plots of $x_2(Y)$ for both equations. The dynamics of the HNLS and ENLS equations are qualitatively similar, but there are $O(1)$ differences in the space/time locations of the fission and fusion events, as well as in the trajectories of the fission products. Again $-L_y\le Y\le L_y$.}\label{NLSMI2_comparison_HNLS_ENLS}
\end{figure}

For completeness, we also present in Figure \ref{trT_cos} the numerical evolution of $\tr T(\lambda_1,Y,t)$, showing that i) it is slowly varying in $Y$; ii) it is essentially constant in time during the linear stages of MI; iii) its variations are around the curves $t_m(Y)$,  in excellent agreement with the theory.
\begin{figure}[h!!!!!!!!]
	\centering
	\includegraphics[width=12cm,height=10cm]{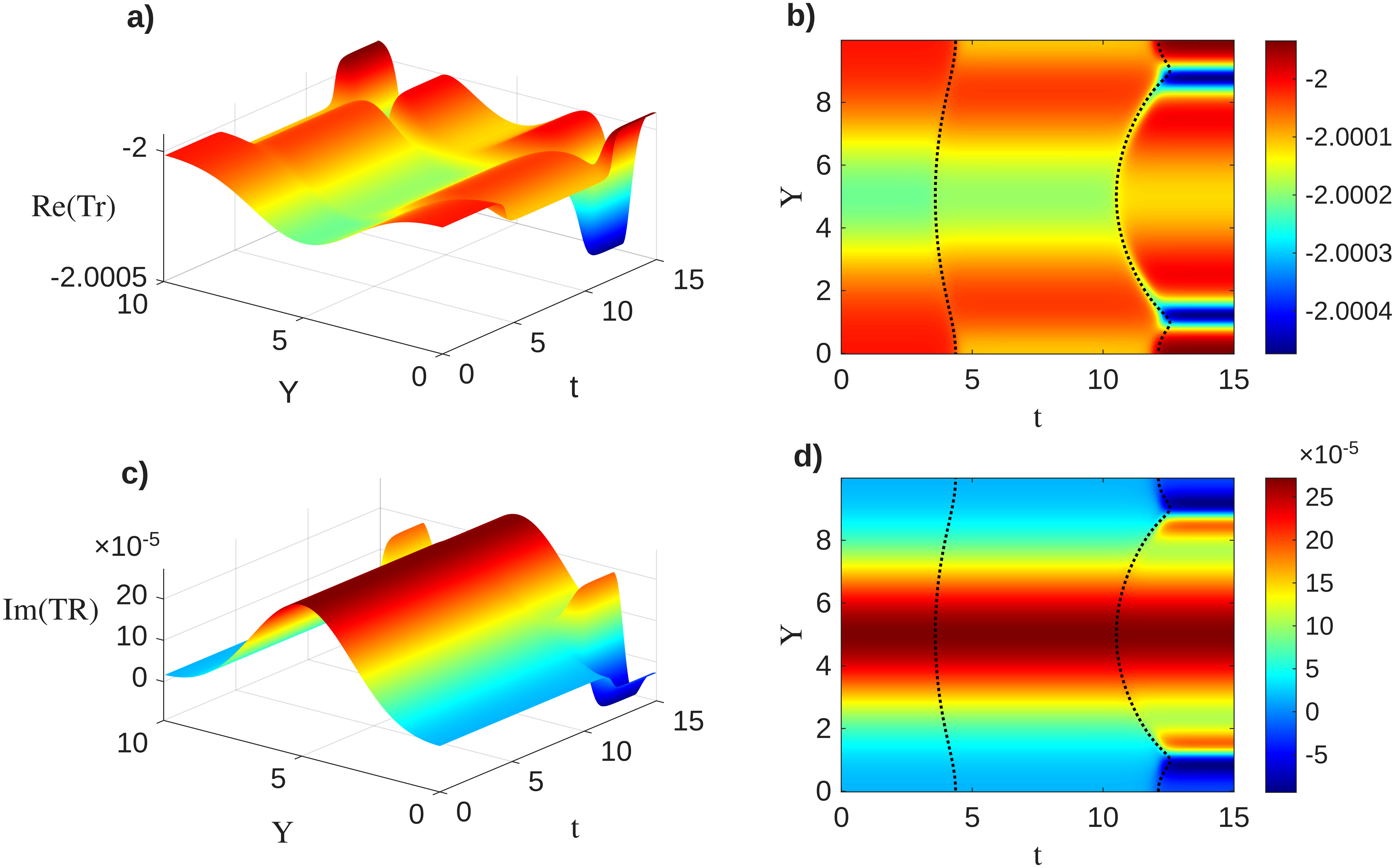}
	\caption{Numerical evolution according to the HNLS equation of the real and imaginary parts of the trace of the monodromy matrix $\tr T(\lambda_1,Y,t)$ over the $(Y,t)$ plane, in a time interval containing the first two nonlinear stages of MI, in excellent agreement with the theory. a) 3D plot of the real part; b) density plot of the real part; c) 3D plot of the imaginary part; d) density plot of the imaginary part. The dashed lines in the density plots are the theoretical curves $t_1(Y)$ and $t_2(Y)$. As predicted by the theory, it is indeed around these curves that Re$(\tr T)$ changes its value, while Im$(\tr T)$ does no vary around $t_1(Y)$.}\label{trT_cos}
\end{figure}

\subsection{The recurrence of $x$-periodic Q1D AWs localized in the $y$ direction over the background}\label{decaying}

Here we make the choice
\beq\label{sechY_Y^2}
f(Y)=\frac{1}{\cosh Y}, \ \ \ g(Y)=d\, Y^2, \ \ d\in\RR,
\eeq
describing a recurrence of Q1D $x$-periodic AWs localized over the background in the transversal $Y$ direction. In the first NLSMI corresponding to \eqref{sechY_Y^2} and investigated in \cite{CS_Q1D_NLSMI1}, the even function $t_1(Y)$ has the global minimum in $Y=0$, with $t_1(Y)=t_{10}+\frac{1}{2\sigma\!_x}Y^2 +O(Y^4)$ for $|Y|\ll 1$, while  $t_1(Y)\sim \frac{1}{\sigma\!_x}(|Y|-\ln 2)+t_{10}$, for $|Y|\gg 1$ (see Figure \ref{sech(Y)_Y^2_t1(Y)_x1(Y)}).
\begin{figure}[h!!!!!!!!]
	\centering
	\includegraphics[width=7cm,height=6cm]{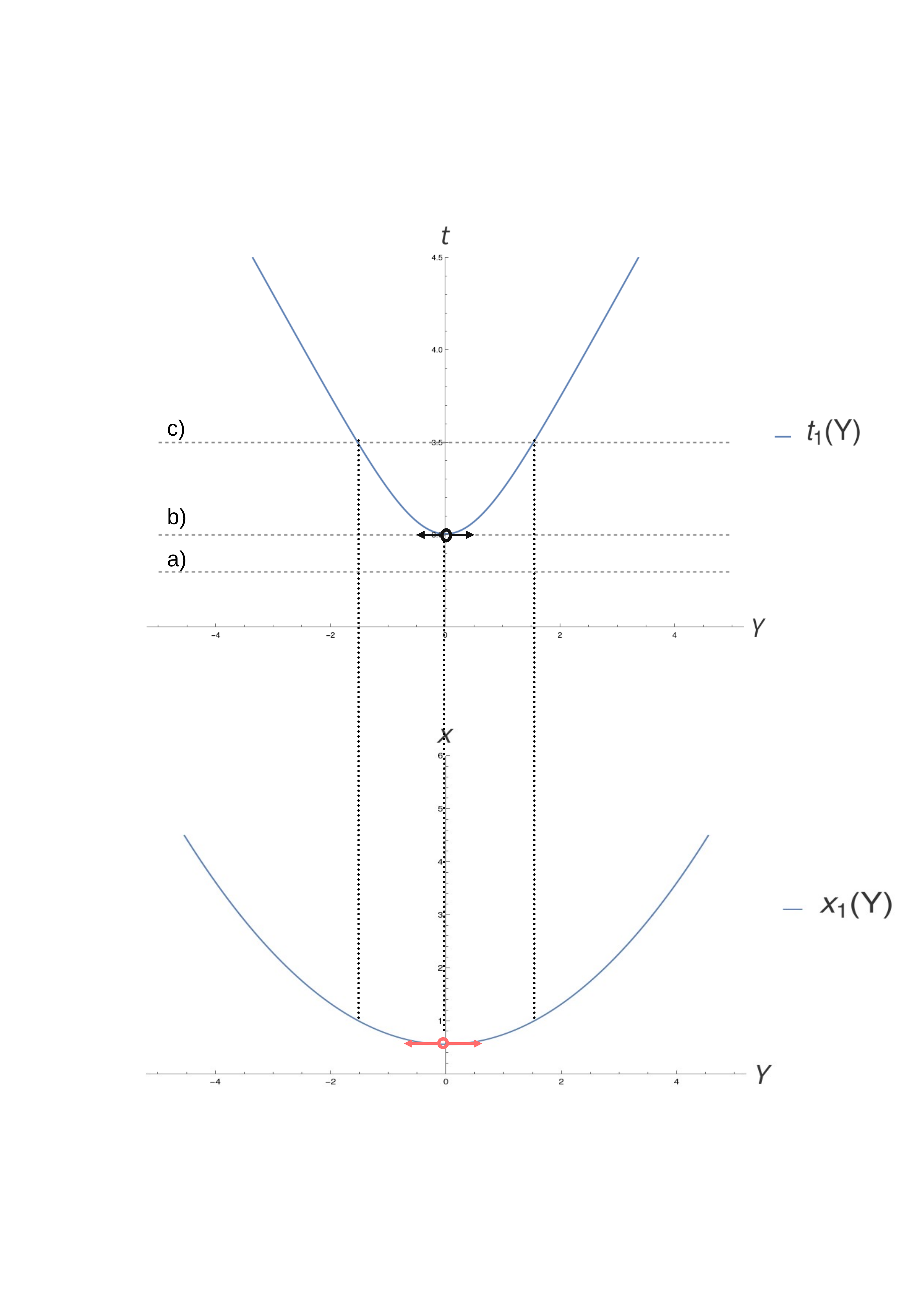}
	\caption{Upper figure: the plot of $t_1(Y)$, together with three horizontal dashed lines at $t=2.8,t_{fiss}=3.005,3.5$ in the $(Y,t)$ plane; lower figure: the parabolic plot of $x_1(Y)$ in the $(Y,x)$ plane. The initial data are $c_{0+}=0.3+i\, 0.2$ and $c_{0-}=i\, 0.5$, with $L_x=6$ and $\eps=0.01$, $\delta=0.001$.}\label{sech(Y)_Y^2_t1(Y)_x1(Y)}
\end{figure}

It follows that the growing AW $u_1(x,Y,t)$ is exponentially localized over the background in the $Y$ direction, undergoing fission at $t=t_{10}$ in $(x,Y)=(x_{10},0)$ with infinite speed at fission time:
\beq
\dot{Y}^{\pm}_m(t)\sim \pm\sqrt{\frac{\sigma\!_x}{2}}\frac{1}{\sqrt{t-t_{10}}}, \ \ 0<t-t_{10}\ll 1,
\eeq
into two waves traveling on the parabolic trajectory $x=x_{1}(Y)$ in opposite direction. For $t-t_{10}\gg 1$ they are well separated and described by
\beq\label{u+-_sech}
  \ba{l}
  u^{\pm}_1(x,Y,t)\sim \frac{\cosh\left\{\left[Y\mp \left(\sigma\!_x(t-t_{10})+\ln 2\right)\mp 2i\phi\right]\right\}+\sin\phi \cos\left[k_x(x-x_{10})-d Y^2\right]}
  {\cosh\left\{\left[Y\mp \left(\sigma\!_x(t-t_{10})+\ln 2\right)\right]\right\}-\sin\phi \cos\left[k_x(x-x_{10})-d Y^2\right]}e^{2it+2i\phi_1},
  \ea
  \eeq
traveling with constant speeds $\pm \sigma\!_x$  and unchanged profile \eqref{u+-_sech} on the trajectories $x=\frac{d}{k_x} Y^2_{\pm}(t)$, where
\beq
\ba{l}
Y_{\pm}(t)\sim \pm\left[\sigma\!_x (t-t_{10})+\ln 2\right], \ \ t-t_{10}\gg 1,
\ea
\eeq
and no fusion takes place (see Figure \ref{sechY_Y^2_3snapshots_NLSMI1}).
\begin{figure}[h!!!!]
  \centering
  \includegraphics[width=12cm,height=4cm]{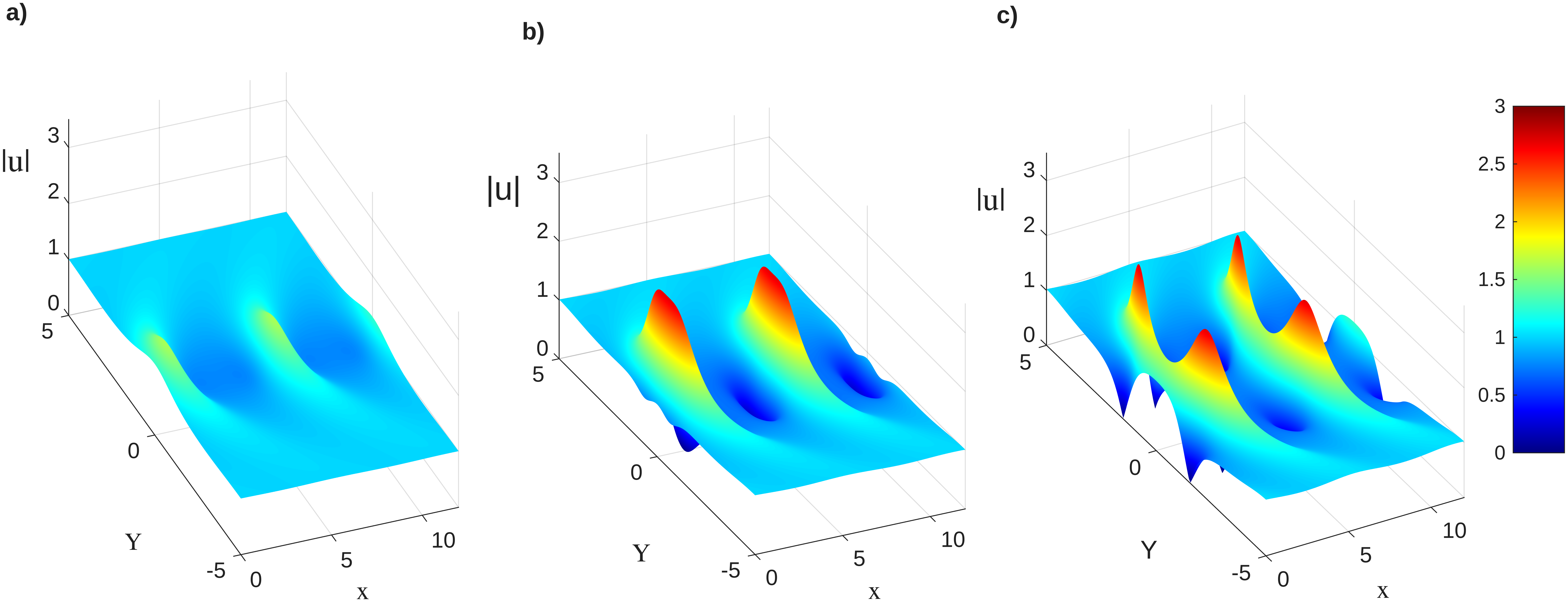}
  \caption{The first NLSMI. 3 snapshots of $|u(x,Y,t)|$ at the time corresponding to the three horizontal dashed lines of Figure \ref{sech(Y)_Y^2_t1(Y)_x1(Y)}. $t=2.8$: the AW growth from the background; $t=t_{fiss}=3.005$: its fission time; $t=3.5$: separation of the fission products. Here we used the same initial data as in Figure \ref{sech(Y)_Y^2_t1(Y)_x1(Y)}, with $-L_x<x<L_x$.}\label{sechY_Y^2_3snapshots_NLSMI1}
      \end{figure}

      The second NLSMI shows richer dynamics with qualitative and quantitative differences between the HNLS and the ENLS equations illustrated in Figure \ref{sech(Y)_Y^2_t2(Y)_x2(Y)}, in which we compare the plots of $t_2(Y),x_2(Y)$ for both equations. While in the HNLS case the even curve $t_2(Y)$ grows monotonically from the global min $Y=0$ to infinity, in the ENLS case, together with the global min $Y=0$, a local max and a local minimum in $Y_{max}$ and $Y_{min}$ are present in between  (see Figures \ref{sech(Y)_Y^2_t2(Y)_x2(Y)}), changing completely the choreography of the second NLSMI (compare Figures \ref{sechY_Y^2_3snapshots_NLSMI2_HNLS} and \ref{sechY_Y^2_7snapshots_NLSMI2_ENLS}).

\begin{figure}[h!!!!!!!!]
  \centering
  \includegraphics[width=7cm,height=10cm]{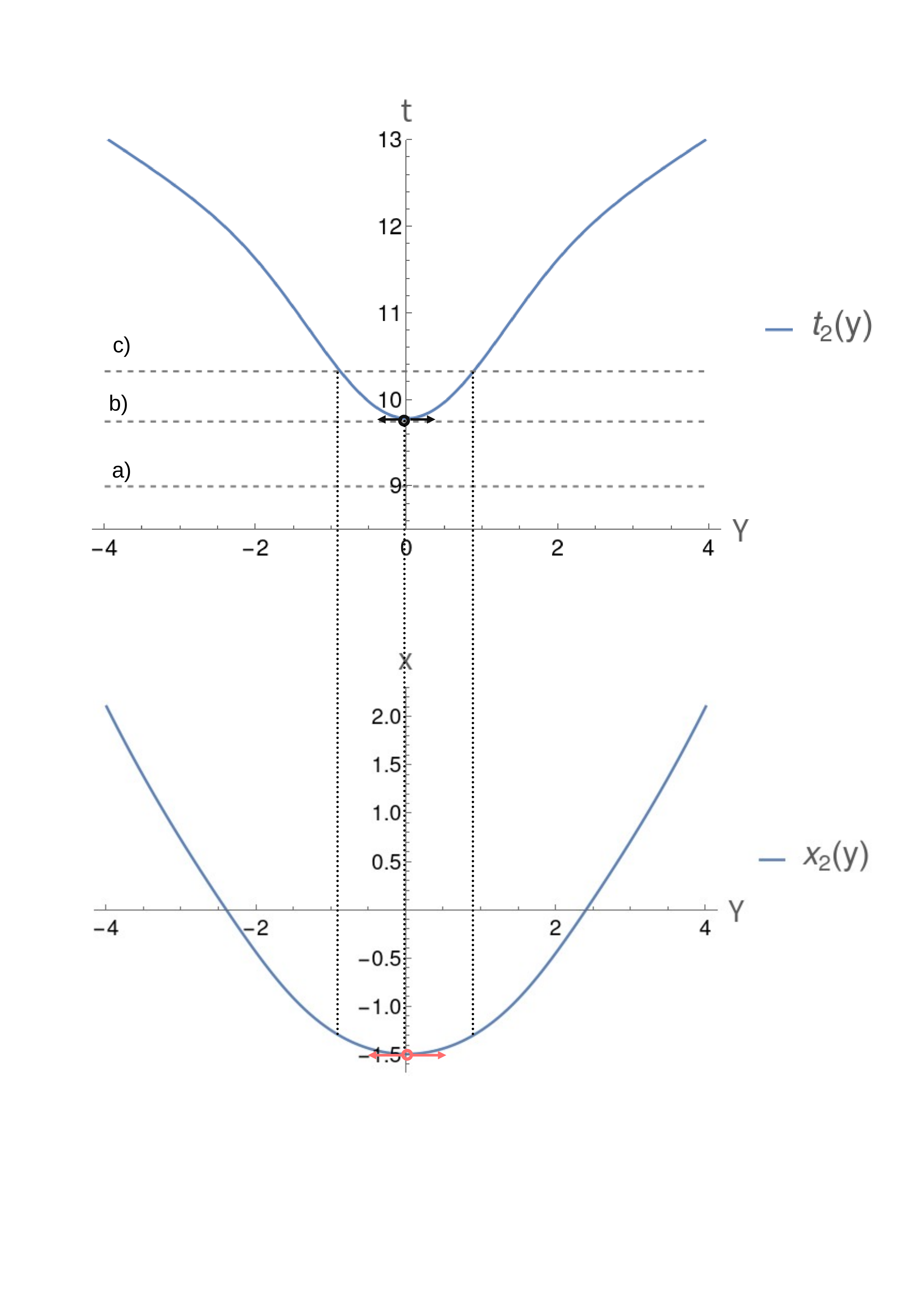}\includegraphics[width=7cm,height=10cm]{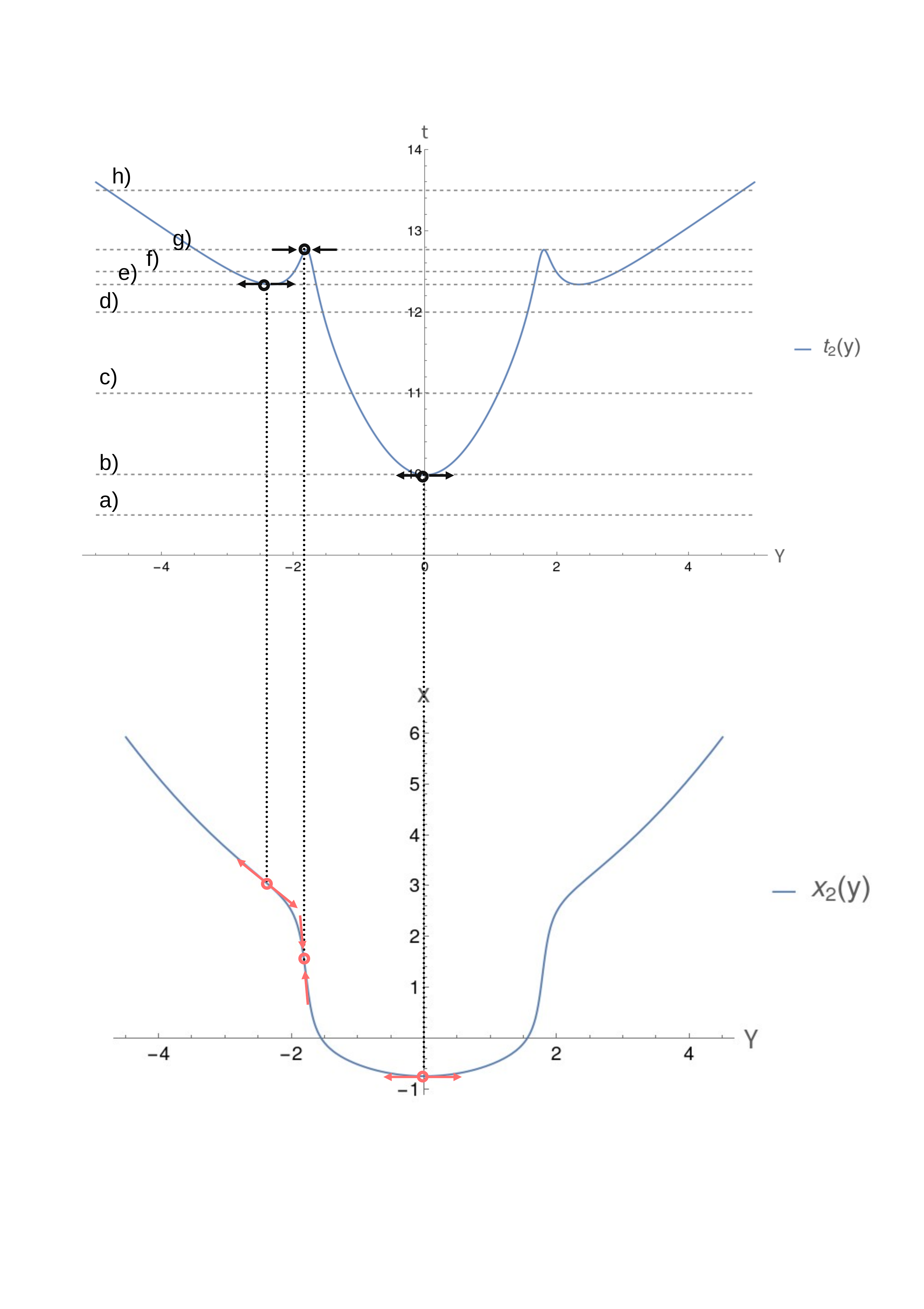}
	\caption{The second NLSMI for both equations. The upper left picture contains the plot of $t_2(Y)$ for the HNLS case, together with the three horizontal dashed lines $t=9.0,t_{fiss}=9.78,10.33$ in the $(Y,t)$ plane, and the lower left picture the plot of $x_2(Y)$ for the HNLS case in the $(Y,x)$ plane. The upper right picture contains the plot of $t_2(Y)$ for the ENLS case, together with the height horizontal dashed lines $t=9.50,t_{fiss}=9.99,11,12,12.34,12.5,12.76,13.5$ in the $(Y,t)$ plane, and the lower right picture the plot of $x_2(Y)$ for the ENLS case in the $(Y,x)$ plane. The initial data are the same as in Figure \ref{sech(Y)_Y^2_t1(Y)_x1(Y)}.}\label{sech(Y)_Y^2_t2(Y)_x2(Y)}
\end{figure}

To be more precise, in the HNLS the AW undergoes fission at $t_2(0)$ in $(x,Y)=(x_2(0),0)$, and the fission products travel in opposite direction along the trajectories $x=x_2(Y)$, mod$L_x$, toward infinity, asymptotically with constant speed $\pm\sigma\!_x$ and unchanged profile; a dynamics qualitatively similar to that of the first NLSMI (see Figure \ref{sechY_Y^2_3snapshots_NLSMI2_HNLS}). 
\begin{figure}[h!!!!]
  \centering
  \includegraphics[width=12cm,height=4.5cm]{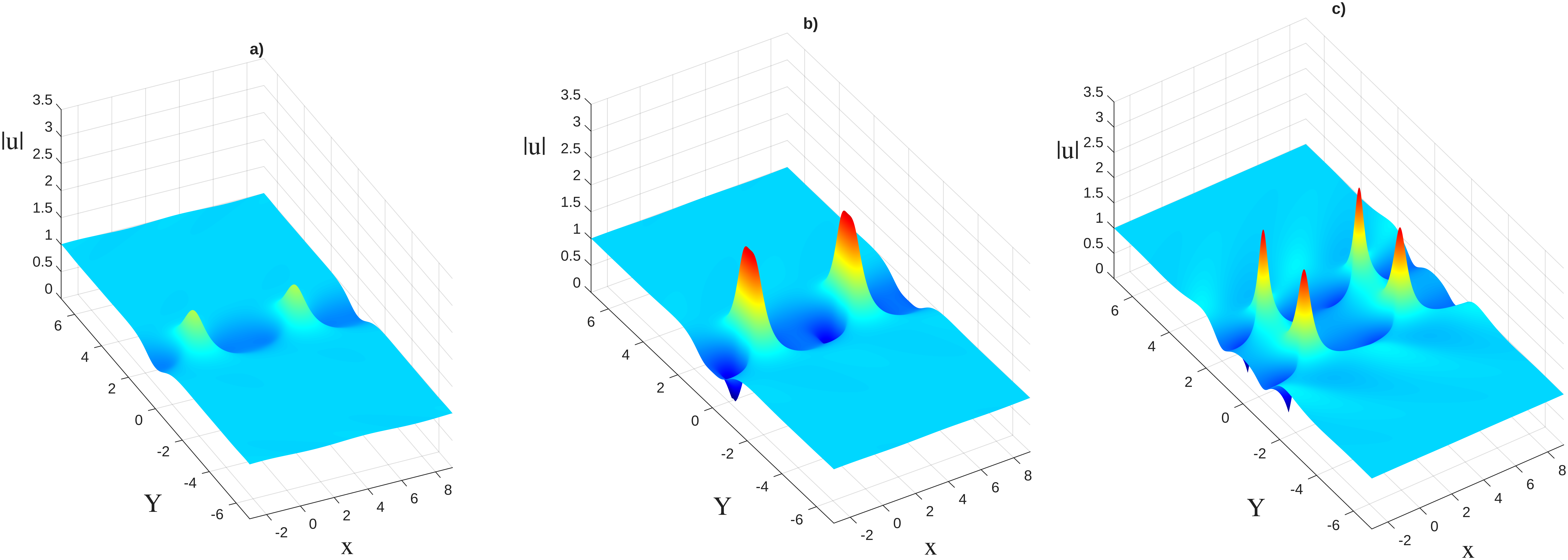}
	\caption{The second NLSMI for the HNLS equation. Three snapshots of $|u(x,Y,t)|$  at the times corresponding to the three horizontal dashed lines of Figure \ref{sech(Y)_Y^2_t2(Y)_x2(Y)}. $t=9.0$: the AW growth from the background; $t=t_{fiss}=9.78$: its fission time; $t=10.33$: separation of the fission products along the parabola $x=x_2(Y)$, mod $L_x$. Here we used the same initial data as in Figure \ref{sech(Y)_Y^2_t1(Y)_x1(Y)}, with $-L_x<x<L_x$.}\label{sechY_Y^2_3snapshots_NLSMI2_HNLS}
      \end{figure}

      In the ENLS case, the (first) AW undergoes fission at $t_2(0)$ in $(x,Y)=(x_2(0),0)$, and the fission products travel in opposite direction along the trajectories $x=x_2(Y)$, mod $L_x$, like in the hyperbolic case, but while these fission products separate, a second AW grows in $(x,Y)=(x_2(Y_{min}),Y_{min})$, undergoing fission at time $t_2(Y_{min})$. Then the fission products of the first fission and the internal fission products of the second fission attract each other along the steepest part of the $x=x_2(Y)$ trajectory, undergoing an oblique fusion at $t_{fus}=t_2(Y_{max})$ in $(x,Y)=(x_2(Y_{max}),Y_{max})$. During this fusion, the two waves exhibit a strong narrowing in the $Y$ direction, and the Q1D regime could be locally lost (the first criticality of this dynamics).  The AW resulting from fusion decays to the background, while the external fission products of the second fission go to infinity without undergoing fusion (see Figure \ref{sechY_Y^2_7snapshots_NLSMI2_ENLS}).

 Since the asymptotics of $t_2(Y)$ are the same for both equations, and those of $x_2(Y)$ are very similar:
\beq
\ba{l}
t_2(Y)\sim \frac{|Y|}{\sigma\!_x}+t_{10}+\ln\left(\frac{(k_x\sigma\!_x)^2}{8\delta^2 |d|}\right), \ \ |Y|\gg 1, \\
x_2(Y)\sim \frac{d}{k_x}Y^2+\frac{\arg(\alpha_{10})}{k_x}+\left(1+\mbox{sign}(b\, d)\right)\frac{L_x}{4}, \ \ |Y|\gg 1, \ \ \mbox{mod }L_x, 
\ea
\eeq
the fission products going to infinity of both equations are described by
\beq\label{u2_fission_prod}
\ba{l}
u_2^{\pm}(x,Y,t)\sim\frac{\cosh\left[ Y\mp Y_{\pm}(t)-2i\phi \right]+\mbox{sign}(b\, d)\sin(\phi) \sin\left[k_x(x-x_{10}-d Y^2) \right]}{\cosh\left[Y\mp Y_{\pm}(t)\right]-\mbox{sign}(b\, d)\sin(\phi) \sin\left[k_x(x-x_{10}-d Y^2) \right]}e^{2it+2i\phi_2}, \\
Y_{\pm}(t)=\sigma\!_x(t-t_{10})+\ln\!\left(\frac{(k_x\sigma\!_x)^2}{8\delta^2 |d|}\right),
\ea
\eeq
traveling with constant speeds $\pm\sigma_x$ and unchanged profile \eqref{u2_fission_prod}, along the asymptotic trajectories $x\sim \frac{d}{k_x}Y^2_{\pm}(t) +\frac{\arg(\alpha_{10})}{k_x}+\left(1+\mbox{sign}(b\, d)\right)\frac{L_x}{4}$, for $t-t_{10}\gg 1$ (see Figure \ref{sechY_Y^2_7snapshots_NLSMI2_ENLS}).

\begin{figure}
  \centering
  \includegraphics[width=13cm,height=8cm]{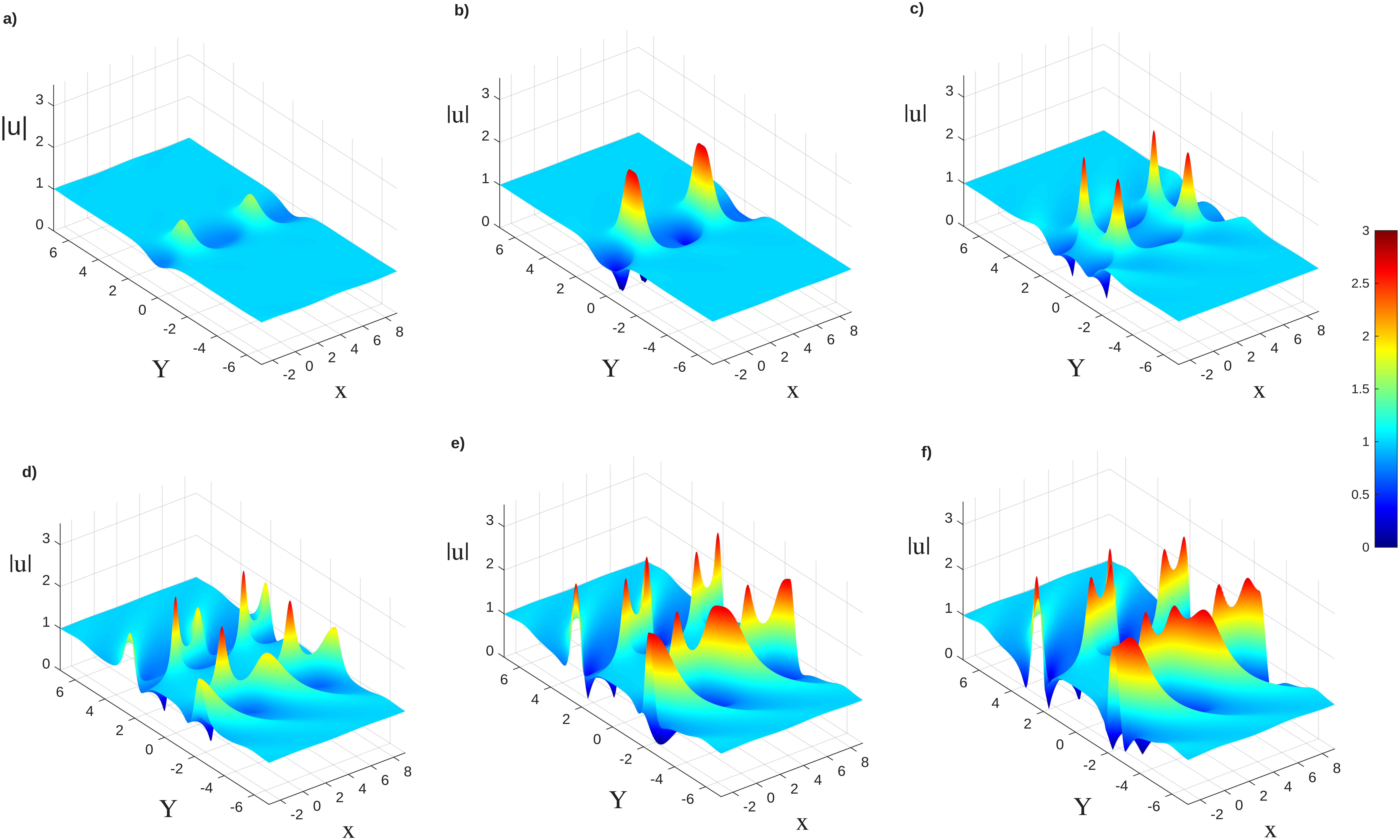}
	\includegraphics[width=8cm,height=3.7cm]{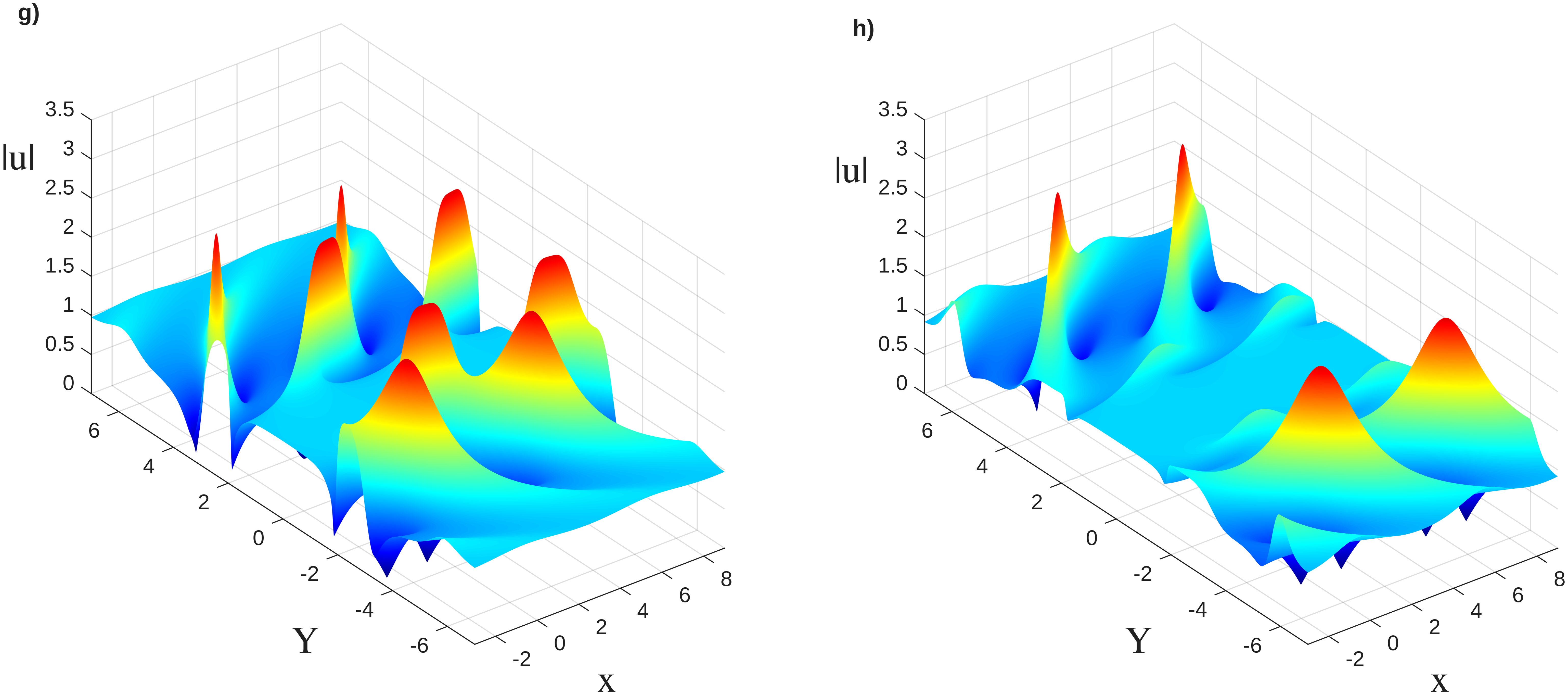}
	\caption{The second NLSMI for the ENLS equation. Height snapshots of $|u(x,Y,t)|$  at times $t=9.50,t_{fiss}=9.99,11,12,12.34,12.5,12.76,13.5$, corresponding respectively to: AW growth; fission time of the first AW; separation of the fission products; while the fission products separate further, a second AW is growing; fission time of the second AW; separation of the fission products of the second AW; oblique fusion of the fission products of the first AW with the internal fission products of the second AW; decay to the background of the AW generated by fusion, while the external fission products of the second AW go to $\infty$ in opposite directions, along the trajectory $x=x_2(Y)$, mod$L_x$. Here we use the same initial data as in  Figure \ref{sech(Y)_Y^2_t1(Y)_x1(Y)}, with $-L_x<x<L_x$.}\label{sechY_Y^2_7snapshots_NLSMI2_ENLS}
      \end{figure}

It is important to remark that, while the time interval of the first NLSMI (AW growth + its fission) is about 1, and the time interval of the subsequent LSMI is about 6, the time interval of the second NLSMI (growth of first AW + its fission + growth of second AW + its fission + fusion and decay) is about 4, invading the time interval of the subsequent LSMI, reduced to $\sim 3$. This is another source of criticality for the analytic formulas of this paper. 

\section{Comparing the theory with the numerical experiments}\label{numerics}

As we did in \cite{CS_Q1D_NLSMI1}, to evaluate how well the leading order analytic formulas \eqref{variation_Qm(Y)}-\eqref{initial} describe the AW recurrence of Q1D AWs, we consider the uniform distance between the numerical evolution, obtained  using the refined $4^{th}$ order Split-step method \cite{Javanainen} and time integration step $dt=10^{-3}$,  and our theoretical approximant as function of time:
\beq\label{uniform_distance}
\| u_{num}-u_{theor}\|_{\infty}(t):=\!\!\!\!\!\!\!\!\!\!\!\!\!\sup_{x\in [0,L_x], Y\in [0,L_Y]}\!\!\!\!\!\!\!\!\!\!\!\!\!\!\!|u_{num}(x,Y,t)-u_{theor}(x,Y,t) |,
\eeq
for both ENLS and HNLS equations, in a time interval containing the first two nonlinear stages of MI.

We first consider the uniform distance \eqref{uniform_distance} during the first two nonlinear stages of MI of the doubly periodic Q1D AW of Section \ref{doubly_periodic}, illustrated in Figures \ref{NLSMI1},\ref{NLSMI2_HNLS}, and \ref{NLSMI2_comparison_HNLS_ENLS}, obtaining an excellent agreement: the two distances are essentially indistiguishable, and their max is less than $2.5\cdot 10^{-3}$ (see Figure \ref{sup_norm}).
\begin{figure}[h!!!!!!!!!!!!!!!!!!!!!!!!!!!!!!]
	\centering
	\includegraphics[trim=0cm 1cm 0cm 0 ,width=11cm,height=5cm]{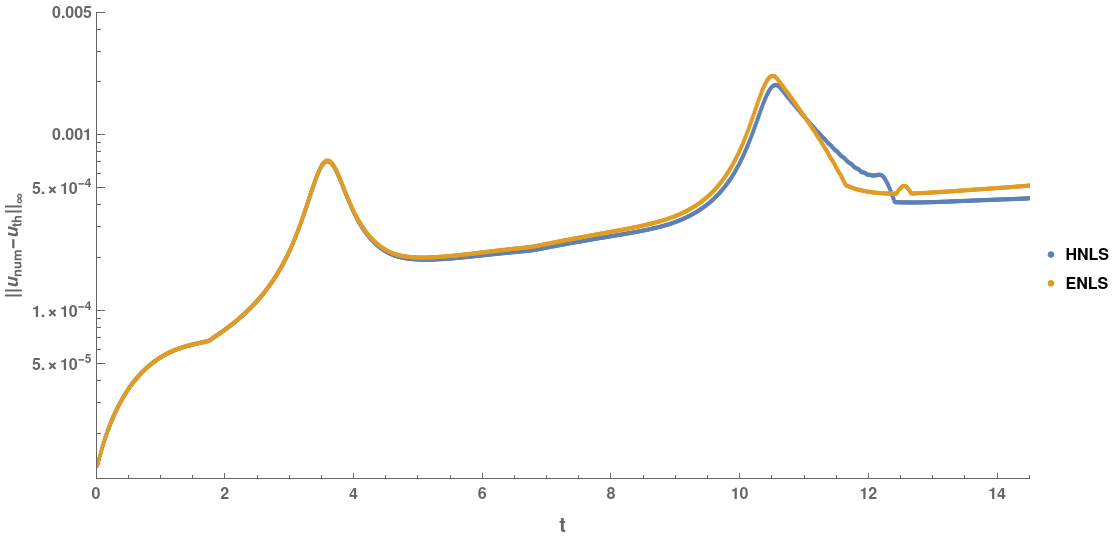}
	\caption{The uniform distance between the numerical output and the theoretical approximant, for both the ENLS and HNLS equations, in a time interval containing the first two nonlinear stages of MI (the two peaks in the figure) described in Figures \ref{NLSMI1} and \ref{NLSMI2_HNLS}, for the doubly periodic AW.}\label{sup_norm}
      \end{figure}

      In the second experiment we consider the uniform distance \eqref{uniform_distance} during the first two nonlinear stages of MI of the dynamics of Section \ref{decaying}, but only for the ENLS equation, since in this case the dynamics is much richer and presents two criticalities absent in the HNLS case: i) a very narrow oblique fusion in which the quasi one dimensionality is clearly in danger, and ii) a temporal duration of the second NLSMI comparable with that of the LSMI preceeding it, and bigger than that of the LSMI following it. Despite these criticalities, the agreement between theory and numerics is very good: only a small oscillation is observed during the oblique fusion, which desappears soon after, and the max of the distance is less than $1.12\cdot 10^{-2}$ (see Figure \ref{sech_norm}).
\begin{figure}[h!!!!!!!!!!!!!!!!!!!!!!!!!!!!!!]
	\centering
	\includegraphics[trim=0cm 1cm 0cm 0 ,width=11cm,height=5cm]{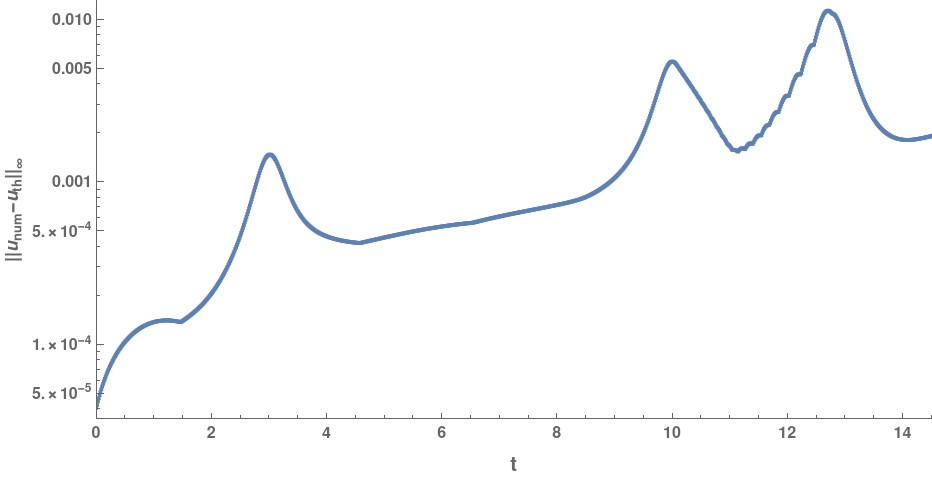}
	\caption{The uniform distance between the numerical output and the theoretical approximant, for ENLS, in a time interval containing the first two nonlinear stages of MI described in Section \ref{decaying}. The first peak corresponds to the single AW appearance in the first NLSMI, and the last two peaks correspond the second NLSMI.  More precisely, the first of the two is associated with the first appearance in the second NLSMI, while the oscillating growth of the third peak corresponds to the very narrow oblique fusion.}\label{sech_norm}
\end{figure}
      
\section{Concluding remarks and future perspectives}\label{Conclusions}

\textbf{R1}. The analytic formulas \eqref{def_um(Y)},\eqref{variation_Qm(Y)},\eqref{variation_xm(Y)_tm(Y)}, and \eqref{initial}, describing at leading order the AW recurrence of Q1D AWs of the $2+1$ dimensional elliptic and hyperbolic NLS equations \eqref{ENLS_HNLS_2+1}, are derived under the following two hypothesis.\\
i) The recurrence time scale $\ln(1/\eps)$ is much smaller than the quasi one dimensionality time scale $1/\delta$: $\ln(1/\eps)\ll 1/\delta$, where $\eps$ and $\delta$ are the two independent small parameters indicating respectively the order of magnitude of the initial perturbation of the background, and the ratio between the $x$-wavelength and the wavelength in the transversal direction. This hypothesis ensures that the term $u_{yy}$ in the MNLS equations remains small for a number of recurrences, before causing blow up in the ENLS case, and $y$-direction defocusing in the HNLS case.\\
ii) Self-consistency: during the evolution described by \eqref{variation_Qm(Y)}-\eqref{initial} the assumptions under which \eqref{variation_Qm(Y)}-\eqref{initial} were derived must continue to be satisfied. Two properties that could cease to be valid during the dynamics are, for instance, ``quasi one dimensionality'', and ``the time interval between two nonlinear stages of MI is logarithmically larger than the time intervals of the nonlinear stages of MI''.
\vskip 5pt
\noindent
\textbf{R2}. Although the analytic formulas \eqref{variation_Qm(Y)}-\eqref{initial} are just the leading order terms of suitable asymptotic expansions, they describe very accurately the real dynamics for a long time, as confirmed by numerous supporting numerical experiments like the ones presented in Section \ref{numerics}. This is the reason why, in this paper, we decided to concentrate on the rich novel features described by them. Rigorous estimates on the errors associated with these leading order formulas, and on the number of recurrences described by them are beyond the scope of this work, and will be presented in a subsequent paper.
\vskip 5pt
\noindent
\textbf{R3}. Increasing $\delta$, the multidimensional perturbation term $u_{yy}$ becomes more and more relevant, and decreasing $\eps$, the MI time scale increases and makes the term $u_{yy}$ acting longer and becoming again more relevant. The quantitative description of both effects is in formula \eqref{variation_Qm(Y)}, in which the two small parameters appear through the combination $(\delta/\eps)^2$. It follows that, no matter how small is $\delta$, if $(\delta/\eps)^2=O(1)$, the second NLSMI and the subsequent ones of the ENLS equation shows $O(|J_m|)$ differences with respect to those of the HNLS equation. Therefore, \textit{while the first NLSMI is essentially the same for both models, and for all the MNLS equations in the Q1D regime, the second NLSMI and the subsequent ones are substantially different for the different MNLS models under scrutiny}. The qualitative explanation of it is the following one: \textit{a very small and difficult to detect difference in the first NLSMI of the ENLS and HNLS equations implies a much bigger difference in the second NLSMI, and in the subsequent ones, due to MI}.
\vskip 5pt
\noindent
\textbf{R4}. For the NLS equation and for its perturbations in $1+1$ dimensions, one deals with a FPUT-type recurrence of the same AB \eqref{Akhmed}, and the $O(1)$ differences among different perturbations of NLS involve only the position and time of the appearances. \textit{In the Q1D theory of MNLS models, one has a recurrence of different adiabatic deformations of the AB, implying that two subsequent nonlinear stages of MI of the same MNLS model exhibit, in general, different combinations of the two basic processes of fission and fusion}.

The results of this paper open two interesting theoretical problems that are presently under investigation by the authors.\\ 
1) the study of the singularities of the Q1D AW recurrence; they arise when, in the iteration,  $Q_{m}(Y)$ possesses zeroes for some $m$, since the zeroes of $Q_{m}(Y)$ are singularities of $t_{m+1}(Y)$, and points in which $x_{m+1}(Y)$ is not well defined (see \eqref{variation_xm(Y)_tm(Y)}). \\ 
2) the study of the fixed points of the  Q1D AW recurrence.
  
The results of this paper have also two natural generalizations we are planning to study: i) to any MNLS equation in $2+1$ dimensions, like the integrable and non integrable Davey Stewartsons equations, and ii) to MNLS equations in $d+1$ dimensions, $d\ge 3$, like the elliptic and hyperbolic NLS equations \eqref{ENLS_HNLS_3+1}.

\vskip 10pt
\noindent
{\bf Acknowledgments}. One of us (PMS) acknowledges useful discussions with G. Biondini. The results of this paper have been obtained in the period 2023-2026. During this period, FC and PMS acknowledge support from the Research Project of National Interest PRIN2020 No. 2020X4T57A, and from the INFN Research Project MMNLP. FC acknowledges support also from PRIN2022 No. 20223T577Z, and from the NSF under grant DMS-2406626.

\end{document}